\documentclass[preprint,12pt]{elsarticle}

\usepackage{amssymb}
\usepackage{amsmath}
\usepackage{url}
\usepackage{framed}
\usepackage{xcolor}
\usepackage{booktabs}
\usepackage{tabularx}
\usepackage{rotating}
\usepackage{float}

\newcommand{\changed}[1]{\textcolor{black}{#1}} %Change blue->black to get a black printout.
\newcommand{\revtwo}[1]{\textcolor{black}{#1}} %Change brown->black to get a black printout.

\journal{Journal of Systems and Software}

\begin{document}

\begin{frontmatter}



\title{Exploring public-private data sharing practices through the lens of Open Data Ecosystems: A multiple case study on municipal and regional freight logistics}

\author[1,2]{John Larsson\corref{cor1}}
\ead{john.larsson@cs.lth.se}
\ead[url]{https://orcid.org/0009-0007-1668-0775}

\author[1,2]{Johan Linåker}
\ead{johan.linaker@ri.se}
\ead[url]{https://orcid.org/0000-0001-9851-1404}

\author[1]{Per Runeson}
\ead{per.runeson@cs.lth.se}
\ead[url]{https://orcid.org/0000-0003-2795-4851}

\cortext[cor1]{Corresponding author.}

\address[1]{Department of Computer Science, Lund University, Lund, Sweden}
\address[2]{RISE Research Institutes of Sweden AB, Lund, Sweden}


%\author{} %% Author name

%\affiliation{organization={},%Department and Organization
            %addressline={}, 
            %city={},
            %postcode={}, 
            %state={},
            %country={}}

%% Abstract
%• Background or Context
%• Objectives or Aims
%• Method
%• Results
%• Conclusions
\begin{abstract}
\textbf{Background:} 
Freight transport is expected to increase, contributing to air pollution. There is, therefore, a critical need for increased efficiency that can reduce emissions. Data sharing is one approach, where actors within Open Data Ecosystems (ODE) collaborate and share data \revtwo{among stakeholders} to plan and operate more efficient and sustainable freight transport.
\textbf{Aim:} 
In this study, we explore how \revtwo{public-private }data sharing practices in municipal and regional last-mile \revtwo{freight} logistics in Sweden may improve efficiency. 
\textbf{Method:}
Following a prolonged participation process, we conducted a multiple-case study involving three different cases, built upon initial meetings, semi-structured interviews, supported by a thematic analysis, and validated through a follow-up workshop. 
\textbf{Results:} \changed{The findings show that data sharing in public-private freight logistics \revtwo{(ODE\textsubscript{PPF})} is shaped by interrelated socio-technical aspects, including platform design}, roles of actors, handling of sensitive data, data quality, and procurement processes. We synthesize our findings in an \revtwo{actor model and a conceptual model}, highlighting  \revtwo{how these aspects contribute to the functioning of ODE\textsubscript{PPF}, and derive seven proposition sketches as a foundation for future research}. 
\textbf{Conclusion:}
Data sharing is a \changed{central mechanism} enabling integration and value creation in last‑mile logistics. \changed{It depends on the interaction between technical and organizational aspects. Public actors can undertake a key role as a neutral keystone actor by orchestrating the ecosystem and facilitating data sharing through governance practices such as procurement.} 

\end{abstract}


% \begin{highlights}
% \end{highlights}

%% Keywords

\end{frontmatter}




\section{Introduction}

In light of the growing freight transport \cite{Trafikverket2023}, air pollution is an increasing challenge \cite{governmentofficesofsweden}, and therefore, finding more efficient freight transport solutions is crucial to achieve national \cite{trafikverket2025,Regeringens_klimathandlingsplan} and global~\cite{SDGS2023} environmental sustainability targets. \changed{One promising approach to addressing these challenges is data sharing between stakeholders, which} enables analysis and planning for more efficient co-loading\revtwo{. We particularly focus on potential emission reduction by} consolidating goods from different logistics actors into a single vehicle \changed{thereby reducing the driven distance and consequently, air pollution}.

\revtwo{Open government data (OGD) refers to public agencies giving access to public data to foster innovation and entrepreneurship~\cite{Lakomaa2013,Dawes2016}. The governance of OGD has been extensively studied, as summarized in several systematic literature reviews, including Attard et al.~\cite{ATTARD2015399} and Wirtz et al. ~\cite{Wirtz2022}, and theorized in detail by Bonina et al.~\cite{Bonina2020}.} 

\revtwo{In contrast, our chosen application domain of municipal and regional freight logistics, embraces the dual relation between public and private partners, where both may function as data supplier and consumer, and where platform orchestration may take different organizational forms. We therefore frame our research in the more general concept of \emph{open data ecosystems} \revtwo{(ODEs)} \cite{Zuiderwijk2014}.}

%The dynamics between actors can be further analyzed through the lens of an 
\revtwo{ODEs are} defined as actors sharing data and collaborating toward a common goal, typically underpinned by a technological platform \cite{Runesonetal2021}. The data sharing can occur at varying levels of openness between involved actors, \revtwo{from Closed, to Shared, to Open} \cite{coyle2020valueofdata}. \changed{While this perspective emphasizes collaboration and coordination among actors, our prior research suggests that realizing such ecosystems in practice remains challenging.} Consequently, data sharing between stakeholders and organizations is not merely a technical challenge, but a complex socio-technical one that covers aspects such as governance, privacy, and legal agreements, and affects multiple business decisions \cite{Jussen2024}. 

\revtwo{In the transportation domain, the research on last mile logistics has grown significantly~\cite{Olsson2019}, particularly as an effect of the increasing number of end-customer deliveries that comes with the growing e-commerce volume. However, interestingly enough, in their systematic literature review of last mile logistics research, Olsson et al. did not find any literature with primary focus on the data aspects of the last mile logistics~\cite{Olsson2019}. We therefore focus on this particular domain from an ODE perspective, considering different levels of openness, referring to the concept as ODE\textsubscript{PPF}, for public-private freight.}

Despite its potential, there are challenges involved in facilitating and implementing this type of data sharing \revtwo{to a new application domain}. These challenges highlight the need to further explore and deepen the understanding of how data sharing is conceptualized \revtwo{in the logistics domain}, and to what extent \revtwo{ODE\textsubscript{PPF}}, capture these issues in practice. To our knowledge, attention has not been paid to the data sharing practices \revtwo{through the lens of ODEs} in the context of public-private partnership in the municipal and regional last-mile freight logistics, constituting the final stage of the delivery process. 

\changed{Building on prior research on \revtwo{ODE} \cite{Zuiderwijk2014, OliveiraMarcelo2019}, related research on data platforms and logistics \cite{OfeDeReuver2025, Moller2020}, as well as our previous literature study \cite{Larsson_Runeson_Linåker_2024}, we explore \revtwo{ODE\textsubscript{PPF}} in public-private partnerships in practice, through three Swedish cases.} In this context, public-private partnerships refer to business relationships between public and private actors engaged in \revtwo{mutual} data sharing \revtwo{in a contractually confined setting}, and to how these shape the functioning of the ecosystem. We also investigate and build upon aspects related to the operations of an \revtwo{ODE\textsubscript{PPF}}, including quality \cite{ATTARD2015399}, legal frameworks~\cite{EU_GDPR_2016}, roles \cite{Zuiderwijk2014, OliveiraMarcelo2019, Immonen2014}, and governance structures \cite{DalBianco2014}. \revtwo{While the technical design of data sharing platforms is a critical enabler for data ecosystems, we find it premature to focus on design theories for the platform before understanding the conceptual ODE\textsubscript{PPF} in the freight logistics context. This study focuses on problem and context conceptualization in design science research \cite{RunesonDSSE2020} and leaves technical solution design for future work.  Such future work should build on existing theory from other domains, for example, Altendeitering et al.~\cite{Altendeitering2024}.} 

This study is a qualitative, multiple-case study where findings are synthesized through a thematic analysis \cite{RunesonEMSE09}. The study follows a prolonged participation process, starting with initial meetings, followed by two iterations of semi-structured interviews. The study includes three cases: Helsingborg Municipality, Södertörn Municipalities, and Predictive Movement. To validate the findings, a follow-up workshop was conducted with the case participants. The study addresses the following research questions: \begin{enumerate}[RQ1] \item What are the conceptual relations and data sharing practices between actors in the last-mile freight transport of goods on behalf of public actors?\item How do actors address quality control, legal compliance, sustainability, governance, and other aspects that they consider important?\item What roles do public actors assume in open data ecosystems? 
\end{enumerate} 

Based on these research questions, this study contributes to \changed{conceptual understanding of} \revtwo{ODE\textsubscript{PPF}} \changed{in the specific domain of municipal and regional freight logistics,} by identifying key components that shape inter-organizational data sharing in public-private contexts. Furthermore, it provides insights into the relationships between \revtwo{actors} involved in data sharing practices, and how these relationships are orchestrated within ecosystems. In addition, the study highlights how key components interact in ecosystems, and illustrates the role of platform configurations in structuring and implementing data sharing in public-private freight ecosystems. \changed{These findings provide a foundation for generalization across the freight domain, contributing to an emerging theoretical understanding of how platform configurations, roles, and governance structures shape data sharing practices in} \revtwo{ODE\textsubscript{PPF}.}

The paper is structured as follows: Section~\ref{sec:background} presents terms, background, and related work. Section~\ref{sec:researchdesign} describes the research design, including a general case description in Section~\ref{sec:cases}. Section~\ref{sec:results} presents the findings, where an actor model and a conceptual model are used to communicate the findings of the study. In Section~\ref{sec:discussion}, we discuss the findings, and Section~\ref{sec:conclusion} concludes the paper. 

\section{Background \& related work}
\label{sec:background}

\subsection{ \revtwo{Actors}}
Table \ref{tab:definition-of-conceptual actors} provides definitions of the conceptual actors that appear in the analysis as well as in the figures. These terms are used to ensure consistency and clarity throughout the study.

\begin{table}[htpb]
\centering
\caption{Definition of conceptual actors recurring in the analysis and related figures.}
\begin{tabularx}{\textwidth}{@{}lX@{}}
\toprule
\textbf{Conceptual actors} & \textbf{Description} \\
\midrule
Public authorities & Actors such as municipalities, governmental authorities.  \\
External consolidation terminal & Operates as a separate unit and is an external component procured by one or more municipalities.  \\
External logistics actor & Actors who provide the logistical services, such as carriers. \\
Supplier of goods & Supplies different actors, such as municipalities with goods. \\
Goods receiver/recipients & The actor who receives the goods from the supplier. \\
Neutral actor/ Non-profit actor & An actor who is considered neutral, and might foster collaboration. \\
\bottomrule
\end{tabularx}
\label{tab:definition-of-conceptual actors}
\end{table}

\subsection{Open data ecosystems}
The definition of \revtwo{ODE} is not universally agreed~\changed{\cite{OliveiraMarcelo2019}. However, as described by Zuiderwijk et al.~\cite{Zuiderwijk2014}, an ODE can be seen as organizations and individuals that collaborate within a network where their common interest forms the ground for the relationships between the actors. Such ecosystems are supported by a technological platform \cite{Jansen2020}, which enables stakeholders in the ODE to process data \cite{OliveiraMarcelo2019}. These capabilities further contribute to innovation, new businesses, and the creation of value. Within ODE, actors collaborate on the data and other resources associated with it, for example, standards or software. The collaboration is facilitated through the exchange of information, resources, and artifacts between different actors across the network \cite{Jansen2020}.} \changed{While embraced in the above definition, Jussen et al.~\cite{Jussen2024} highlight that data-sharing in ecosystems has complex socio-technical challenges, including issues related to governance, contractual agreements, and data quality. These challenges are evident from the need to coordinate multiple actors, combining roles and responsibilities, and establish common standards between data providers and consumers sharing data on a platform.} 

\changed{In relation to the overall aim of this study to reduce air pollution in freight logistics, Melville et al.~\cite{Melville2017} indicate that data and information systems play a significant role in enabling reduced environmental impact by supporting more informed decision-making and low-carbon operations. The study also emphasizes that systems, data accuracy, and completeness are critical for enabling efficient decision-making aimed at reducing emissions and achieving low-carbon operations~\cite{Melville2017}.}

\subsection{Activities, elements, and \revtwo{actors} within \revtwo{open data ecosystems}}
\revtwo{Building on the definition of ODEs presented in the previous section, Zuiderwijk et al.~\cite{Zuiderwijk2014} identify several activities and elements within the ODEs, involving} different \revtwo{actors} (e.g., user and producer of data). \revtwo{These activities include} \textit{searching, cleansing, viewing, requesting, analyzing, and combining data.} Furthermore, they describe that previous research had a primary focus on fewer activities, and not on the larger picture of ODE, and Zuiderwijk et al.~\cite{Zuiderwijk2014} therefore have chosen to highlight several interlinked key elements that an ODE should address. The key elements are: \textit{publish data, explore data and their related licenses, analysis of data, interpret the data, as well as giving feedback to the producer.} Several other elements are also described as essential for ODE, \textit{user pathways for the data, quality management system, and metadata for connection between the elements.} These last elements should then act as an enabler for interoperability between the elements of the ecosystem.

Multiple roles within ODE are also found in the literature, where Oliveira et al.~\cite{OliveiraMarcelo2019} and Zuiderwijk et al.~\cite{Zuiderwijk2014} highlight the \revtwo{actors} of the data provider, intermediaries, and data user. Furthermore, Oliveira et al. and Immonen et al.~\cite{Immonen2014}, emphasize that there are multiple other \revtwo{actors} connected to ecosystems, while Oliveira et al. also describe that some \revtwo{actors} can represent the same duty. An example is when data providers and intermediaries produce data. It can be noted that data providers are not required to have the role of data producer, but \revtwo{the role} can be carried out by a \revtwo{separate actor within} the ecosystem. 

\changed{In ecosystem research, Jacobides et al.~\cite{Jacobides2018} further describe ecosystems as sets of interacting actors linked through complementarities, where complementary products and services jointly contribute to value creation in the ecosystem. These complementarities exist at the level of roles, linking different types of actors, such as complementors and suppliers. Furthermore, as Adner \cite{Adner2017} highlights, ecosystems consist of multiple stakeholders that need to coordinate to create value propositions to materialize.}

\subsection{Governance structure of \revtwo{open data ecosystems}}
\label{sec:Governance structure of ODE.}
Data sharing ecosystems may be structured in different ways, \changed{as proposed by Dal Bianco et al.~\cite{DalBianco2014}, ranging from \textit{organization-centric, to consortium-based, and community-based}. We consider these structures} to have different meanings for the ecosystem: organization-centric governance is based on a single actor. Consortium-based governance \revtwo{is jointly managed by} actors or organizations, which may also be co-owned by the ecosystem actors. Community-based ecosystems have a different type of setup, where the governance might be structured around the actors who are members of the ecosystem. 

\revtwo{Bartolomucci et al.~\cite{Bartolomucci2026} highlight the central role of governance in data-sharing ecosystem,} \changed{including dependency on both collaborative and data governance dimensions. The governance dimensions should be viewed together from a holistic socio-technical perspective, and their integration also influences how value between these socio-technical aspects is derived from data. The relationship between these dimensions further builds the foundation for effectiveness, regulating how data sharing is enabled and maintained.}

\changed{A keystone actor might also be found in the ecosystem, as described by Oliveira et al \cite{OliveiraMarcelo2019}}. Jansen \cite{Jansen_2014}, and Harrison et al.~\cite{Harrison2012} further describe the keystone actor, where this actor can have an impact on the health of the ecosystem. Regarding the structure of the ecosystem, Oliveira et al.~\cite{OliveiraMarcelo2019} describe this in terms of five structures: \textit{keystone-centric, business model-oriented, marketplace-based, platform-centric, and intermediary-based.} As shown in the study by Oliveira, keystone-centric can be described as a structure where actors in the ecosystem are organized around the keystone. It is also explained by Oliveira and Lóscio \cite{Oliveira_Lóscio_2018}, that a keystone-centric structure can be a government administration or agency, responsible for providing a greater amount of data within the ecosystem, but with limited control over other actors, \changed{following the structure of the governance model by Nakakoji et al.~\cite{Nakakojietal2002}.}

\subsection{Quality of data in \revtwo{open data ecosystems}}
\label{sec:DataQuality}
\changed{Strong et al.~\cite{Strongetal1997} describe \revtwo{data} quality dimensions as accuracy, accessibility, relevance, completeness, and interpretability. Building further on these dimensions \cite{ATTARD2015399, VETRO2016325, Slibar2021, Moradi2022}, data quality in ODE can be understood through several key dimensions, including usability, understandability, accuracy, timeliness, completeness, openness, consistency, and compliance.
\revtwo{Prior research} also highlights that \revtwo{low data quality can hinder data sharing \cite{Bartolomucci2026} and} limit} the ecosystem \revtwo{from} reaching its full potential. 

\subsection{Legal aspects of \revtwo{open data ecosystems}}
\revtwo{Legal aspects related to data sharing include legislation (such as the EU General Data Protection Regulation (GDPR) \cite{EU_GDPR_2016}) and licensing.} Connected to data sharing, there is an uncertainty around legal complications where Soosaar and Nikiforova \cite{Rajamäe_Soosaar_Nikiforova_2025}, describe how municipalities can feel fearful of not complying with data protection regulations. However, they also describe that an alternative to avoid this can be to implement guidelines and processes to comply with the laws that exist, such as GDPR. 

Regarding licensing, there is an additional complexity when data is shared. If an organization chooses to share data, liability may remain with the original data-providing organization, what then governs is the type of user agreement and license chosen to use. Solutions for the legal uncertainties are also discussed in the literature, where clearer guidelines and frameworks \cite{Dulong_de_Rosnay_Janssen_2014}, \cite{Magalhäs2016}, and agreements may be alternatives.

\subsection{Data sharing platform}
\changed{Data platforms enable actors to interact and share data, where openness and interoperability have important roles in reducing restrictions and allowing stakeholders to access the platform. Platform openness is also considered a key aspect, allowing actors to use artifacts and reach beyond the local boundaries of the platform \cite{OfeDeReuver2025}.  In the context of logistics, Möller et al.~\cite{Moller2020}, highlight how digital platforms are related to data-driven business models and services, viewing data as a key resource. The data plays an important role in visibility and optimization services, which support transparency and data-driven decision-making, including logistics planning.} \revtwo{More generally, Altendeitering et al. \cite{Altendeitering2024} focus on data quality and provide prescriptive design theory for data quality tools in data ecosystems, emerging from three industrial cases. With respect to interplay in data ecosystems,} Pan et al.~\cite{PAN_et_al_2021} describe the importance of digital interoperability between systems and involved actors. They also put forward four major challenges related to digital interoperability: \textit{Data sharing formats, communication, security, and type of data needed.} Furthermore, there are various proposed solutions to these key challenges, with \textit{smart infrastructure and communities} being one solution. This solution highlights how data sharing and interoperability are central \changed{not only to effective logistics solutions but also} to activities such as efficient planning and communication between actors.

\section{Research design}
\label{sec:researchdesign}
\begin{figure}[t]
    \centering
    \includegraphics[width=0.95\linewidth]{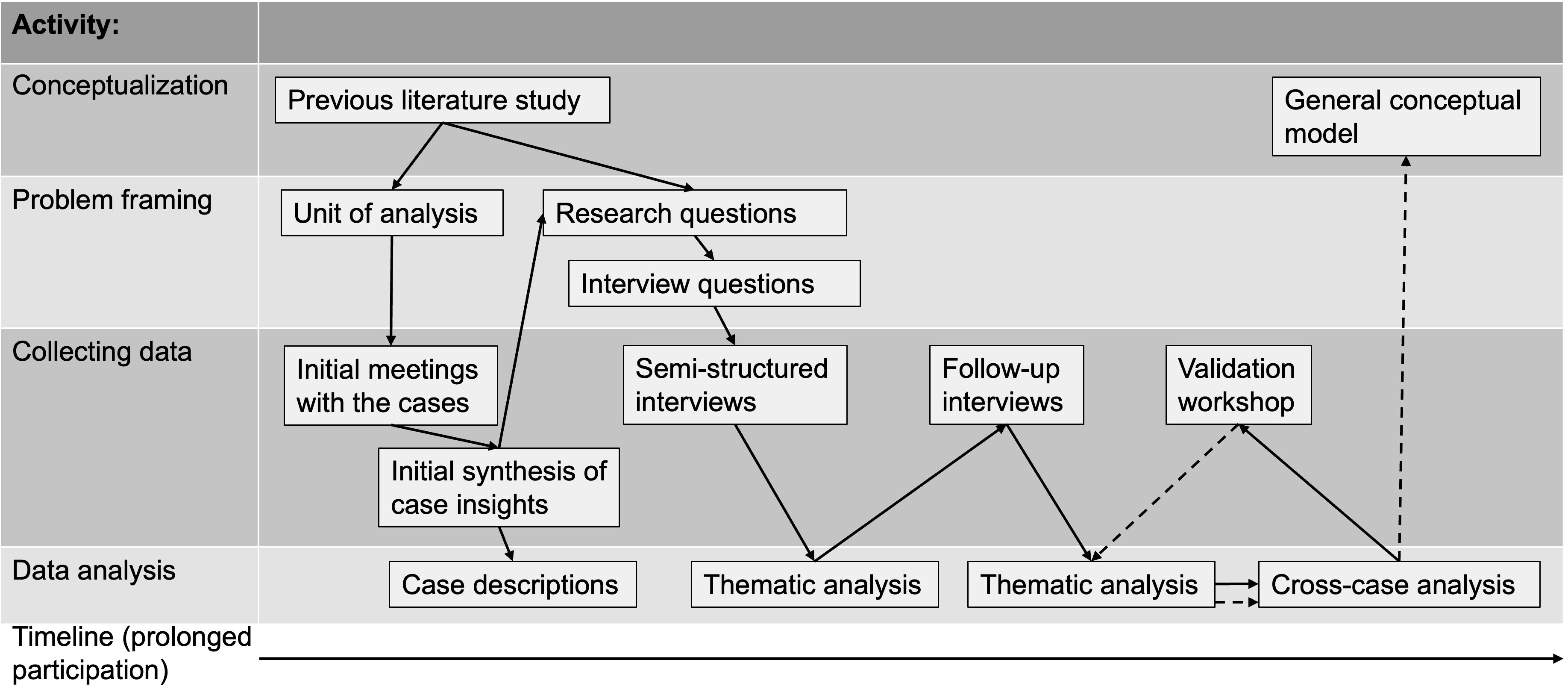}
    \caption{Overview of the methodological approach, describing the different activities involved in the prolonged participation. White boxes represent activities, and gray lanes represent different types of research activity. The timeline begins with our previous literature study \cite{Larsson_Runeson_Linåker_2024} and concludes with the development of the conceptual model \revtwo{and the propositions}, capturing the core of our findings. The solid arrows indicate how the activities are connected over time, and the dashed arrows indicate the progress of data collection and analysis from the validation workshop.}
    \label{fig:Research design}
\end{figure}

\revtwo{This study contributes to the problem conceptualization within a scope of design science research \cite{RunesonDSSE2020}, aiming to provide design knowledge for freight logistics governance and collaboration platforms. For this purpose}, we utilized a multiple case study methodology approach~\cite{Casestudy2012} \changed{and designed a qualitative multiple-case study to facilitate deeper understanding of data sharing in practice within specific contexts. The cases were strategically selected to capture variation in public-private data sharing initiatives within municipal last-mile freight logistics. The three cases --} Helsingborg Municipality, Södertörn Municipalities, and Predictive Movement, which is orchestrated by Luleå University of Technology \changed{-- represent different configurations of data sharing initiatives, enabling analytical comparison across governance structures, levels of maturity, and platform configurations}\revtwo{, in line with Flyvbjerg's \emph{maximum variation} case selection strategy~\cite{Flyvbjerg2006}.}
A brief overview of the flow of research activities is provided in Figure \ref{fig:Research design}. 

\subsection{Unit of analysis}
The unit of analysis for this study is the participation of municipal and public actors, in \revtwo{handling shared} open data within freight transport on behalf of public actors, with the aim of increasing the efficiency of municipal last-mile logistics. The study explores various roles, such as data consumers, producers, and platform providers. The study also explores how these actors address quality, legal, and governance aspects. The selection of the unit of analysis is inspired by the findings of our previous literature study \cite{Larsson_Runeson_Linåker_2024}, which is illustrated in Figure \ref{fig:Research design}, acting as a starting point for the study.

\subsection{Data collection}
\label{sec:datacollection}
The primary empirical data collection is interview-based, with interactions with interviewees taking different forms over time. \revtwo{Field notes from} prolonged participation \cite{Casestudy2012} served as the basis for the interaction between researchers and actors within the cases. 
This is illustrated in the data collection lane in Figure~\ref{fig:Research design}, where the exploration of the cases began with initial meetings with respondents from each case. The initial meetings allowed the researchers to take field notes, which were subsequently used to develop initial syntheses. These syntheses were updated after each meeting, allowing for knowledge development, and served as the foundation for the case descriptions in Section~\ref{sec:cases}. In the case of Helsingborg Municipality, five respondents participated, in Södertörn, two, and in the case of Predictive Movement, three respondents, as reported in Table \ref{tab:respondents_participation}. \revtwo{The limited number of respondents in Södertörn Municipalities reflects that only two relevant persons agreed to participate in the study. Both respondents were considered relevant to the study due to their expertise and organizational role in procurement and logistics operations.} After initial meetings with each case, the researchers conducted two rounds of semi-structured interviews with respondents from the cases. The semi-structured interviews involved nine respondents, five of whom had also participated in the initial meetings, while four additional respondents were considered relevant for the interview phase based on snowball sampling \cite{Wohlin2014} within the cases. 

The nine respondents participating in the first interview rounds possess backgrounds in both public and private sector organizations, as well as in academia. An overview of the respondents' roles, years of professional experience, and participation across interview rounds and the validating workshop is presented in Table~\ref{tab:respondents_participation}. The table shows the participants’ professional roles, their years of experience in the field, and their respondent codes: HBG1–HBG4 refer to Helsingborg Municipality, and PF1 refers to a private company connected to one of the cases. ST1–ST2 refer to Södertörn, and PM1–PM2 refer to the respondents from Predictive Movement. These acronyms are used throughout the results to support the transparency and traceability of the findings. Furthermore, HBG5-HBG7 and PM3-PM4 refer to the additional respondents from the initial interviews. 

\changed{The respondents comprise a combination of technical and domain expertise. Some participants had strong technical backgrounds and were involved in the design and development process of data sharing platforms and Transport Management Systems (TMS). Other respondents held organizational roles related to procurement and logistics planning, contributing with operational insights into data sharing practices. Together, this ensured that both technical and organizational perspectives were captured.}

\begin{table}[t]
\centering
\caption{\changed{Overview of respondents, their roles, professional experience, and participation across data collection activities, Interview (I) rounds 1-3, and validating Workshop (W).}}
\begin{tabular}{lp{4.5cm}rcccc}
\hline
\textbf{ID} & \textbf{Role} & \textbf{Years} & \textbf{I1} & \textbf{I2} & \textbf{I3} & \textbf{W} \\
\hline
HBG1 & Traffic Planner & 14 & X & X & X & X \\
HBG2 & Solutions Architect & 4 &  & X &  &  \\
HBG3 & User Experience Designer & 6 & X & X &  & X \\
HBG4 & Sustainability Strategist & 9 &  & X &  &  \\
HBG5 & Procurement Manager & 5 & X &  &  &  \\ 
HBG6 & Innovation Lead & 5 & X &  &  &  \\ 
HBG7 & Innovation Catalyst & 3 & X &  &  &  \\
PF1  & Transport Manager & 6 &  & X &  &  \\
ST1  & Procurement Logistics Specialist & 8 & X & X & X & X \\
ST2  & Coordinator Goods Distribution & 5 & X & X &  &  \\
PM1  & University Lecturer & 13 & X & X &  & X \\
PM2  & Postdoctoral Researcher & 5 &  & X &  &  \\
PM3  & Project Manager & 6 & X &  &  &  \\
PM4  & Associate Professor & 2 & X &  &  &  \\
\hline
\end{tabular}
\label{tab:respondents_participation}
\end{table}

The first round of semi-structured interviews closely followed the interview questions (see~\ref{app1}), and included participants are illustrated in Table \ref{tab:respondents_participation}, column I1. The interview questions were sent to the respondents well in advance, allowing them to prepare their answers. As illustrated in Figure \ref{fig:Research design}, the research questions that formed the basis for the interview questions were derived from our previous study \cite{Larsson_Runeson_Linåker_2024}, which examined the theoretical foundations. 

Following the prolonged participation, a second round of follow-up interviews was conducted \revtwo{with respondents from all three cases, as well as an external actor,} as shown in Table~\ref{tab:respondents_participation}, \revtwo{column I2. To further enhance our understanding, a third round of interviews was conducted with respondents from Helsingborg Municipality and Södertörn Municipalities, as shown in Table~\ref{tab:respondents_participation}, column I3.}

After conducting the interviews, % several tools were employed for transcribing, coding, and conducting a thematic analysis. All interviews conducted were 
the recorded voices were transcribed using the open source tool WhisperX\footnote{https://github.com/m-bain/whisperX}. %, which facilitated automation. 
The raw text data generated by WhisperX was thoroughly reviewed by the researchers to ensure its accuracy. During the coding and thematic analysis process, a spreadsheet was used to record the process. Also, a codebook was created in the spreadsheet (presented in Section~\ref{sec:results}, Table \ref{tab:themes}) defining the codes and themes used to describe the data. To validate the findings, a workshop was held with respondents from the different cases, as illustrated in Figure \ref{fig:Research design}. As shown in Table \ref{tab:respondents_participation}, \revtwo{column W}, two respondents from Helsingborg Municipality, one from Södertörn Municipalities, and one from Predictive Movement participated in the workshop. The workshop was recorded and transcribed, and as the figure shows, the collected data were also analyzed and used in the thematic analysis, before contributing to the development of the General actor model (Section \ref{sec:results}, Figure \ref{fig:conceptual}).

\subsection{Thematic analysis}

For the analysis of data, we conducted a thematic analysis inspired by Braun and Clarke \cite{BraunClarke2006}. The thematic analysis followed a deductive thematic approach, \changed{guided by concepts derived from our prior research on data sharing in public-sector logistics \cite{Larsson_Runeson_Linåker_2024}. More specifically, the analysis focused on concepts related to data sharing practices, governance structures, data quality, sensitive data, and sustainability aspects.} \changed{The analysis followed Braun and Clarke's six-step guide \cite{BraunClarke2006}, including familiarization with the data, initial coding, theme identification, and evaluating and refining the themes. For the analysis, the raw data were categorized into themes using codes based on the study's research questions. To conduct this analysis, a structured coding spreadsheet was used, where the raw data were both color-coded and labeled with text-based codes. The codes were then organized into themes based on the study’s research questions and unit of analysis.}

\revtwo{The codes applied and generated through the deductive approach resulted in seven distinct themes in Table \ref{tab:themes}, which is a summary of our codebook. These themes also served as the basis for the subthemes, which are visualized in the conceptual model in Figure \ref{fig:Analysis model}, describing how our findings emerged from the empirical observations. In this model, we traced how each theme %from Table \ref{tab:themes} 
leads to its corresponding subthemes and their relation to the research questions.
From the individual case descriptions, we also derived a joint actor model, see
Figure~\ref{fig:conceptual}}. \changed{To enhance analytical rigor, the coding process included discussion and comparison among the authors, allowing for evaluation and refinement of the themes across the cases.} \changed{In addition, we deduced proposition sketches as a first step towards theorization, following the principles of Sjøberg et al.~\cite{Sjberg2008}. }

\subsection{Ethical considerations}
Several ethical aspects have been considered when conducting this study,
inspired by Strandberg \cite{StrandbergEthical2019}. These include informed consent, confidentiality, and secure storage of data. Before conducting interviews with the selected respondents, they were asked for permission to participate in the study, consent to the recording of the interview, and agreement for the researcher to transcribe and store the data on a secure server. The transcribed raw data from the interviews were also returned to each respondent, allowing them to review the material and make changes or removals. Furthermore, to safeguard respondent confidentiality, each respondent was assigned an anonymity code %. These codes are presented in Section~\ref{sec:datacollection}, 
(Table~\ref{tab:respondents_participation}).

\subsection{Threats to validity}
This section outlines threats to validity related to the study and focuses on internal validity, external validity, reliability, and construct validity \cite{Casestudy2012}. In addition, we present approaches employed to mitigate the identified threats.  

\subsubsection{Internal validity}
A thematic analysis with deductive coding formed the basis for structuring and analyzing the raw data collected from the respondents. This, in turn, can contribute to researcher bias as there is a risk that the data are interpreted based on previous research, which can influence the themes selected. To minimize the impact of this, the researchers have employed several strategies to strengthen internal validity: The initial coding of the data was performed by the first author. The second and third authors then independently coded a subset of the data to validate the coding framework. The results were then compared and discussed to align interpretations and strengthen the robustness of the thematic analysis. A validation workshop was held with the respondents, in which findings were presented. Participants were invited to provide feedback on the collected data, and this process supported internal validity by allowing respondents to validate the interpretations and ensure that the findings reflected their experiences.

\subsubsection{External validity}
To strengthen the external validity of the study, the three cases included in the study are from different geographical areas of Sweden. However, the limited sample size might restrict transferability to other contexts, \revtwo{while the  uneven number of respondents across the cases may affect the depth and nuance of analysis across the three cases}. Though some results might be transferable within Sweden, this limitation should be taken into consideration when evaluating the findings. 

\subsubsection{Reliability}
Several technical tools were used to transcribe, code, and conduct a thematic analysis in order to strengthen the reliability. However, there is still a risk of misinterpretation in the creation of a thematic analysis, and this should therefore be taken into consideration.

\subsubsection{Construct validity }
To strengthen construct validity in this study, the researchers aimed to ensure that the interview questions were closely aligned with the theoretical constructs being investigated. This alignment was intended to ensure that the collected data directly reflect constructs such as data sharing, data quality, legal aspects, and governance structures. Additionally, the validation of the coding process by the second and third authors was established to ensure that the themes identified in the analysis accurately represented the intended constructs, rather than individual researcher bias.

\changed{\subsection{Case descriptions}
\label{sec:cases}}
This section describes the three explored cases. The descriptions of each case are derived from the initial meetings, explained in Section~\ref{sec:datacollection}, and illustrated in Figure \ref{fig:Research design}. Each subsection outlines key characteristics of the cases considered relevant for the following analysis and presents \revtwo{an actor model} for each case.
 
\changed{\subsubsection{Helsingborg Municipality}}

Helsingborg Municipality, with approximately 150,000 inhabitants, has implemented a digital system for monitoring freight transport connected to the municipality's organizations, such as schools and public institutions. The digital system, the moderately mature Sam-application, is a mobile platform that utilizes geofencing technology which enables data sharing between involved actors. 

\begin{figure} [htbp]
    \centering
    \includegraphics[width=0.75\linewidth]{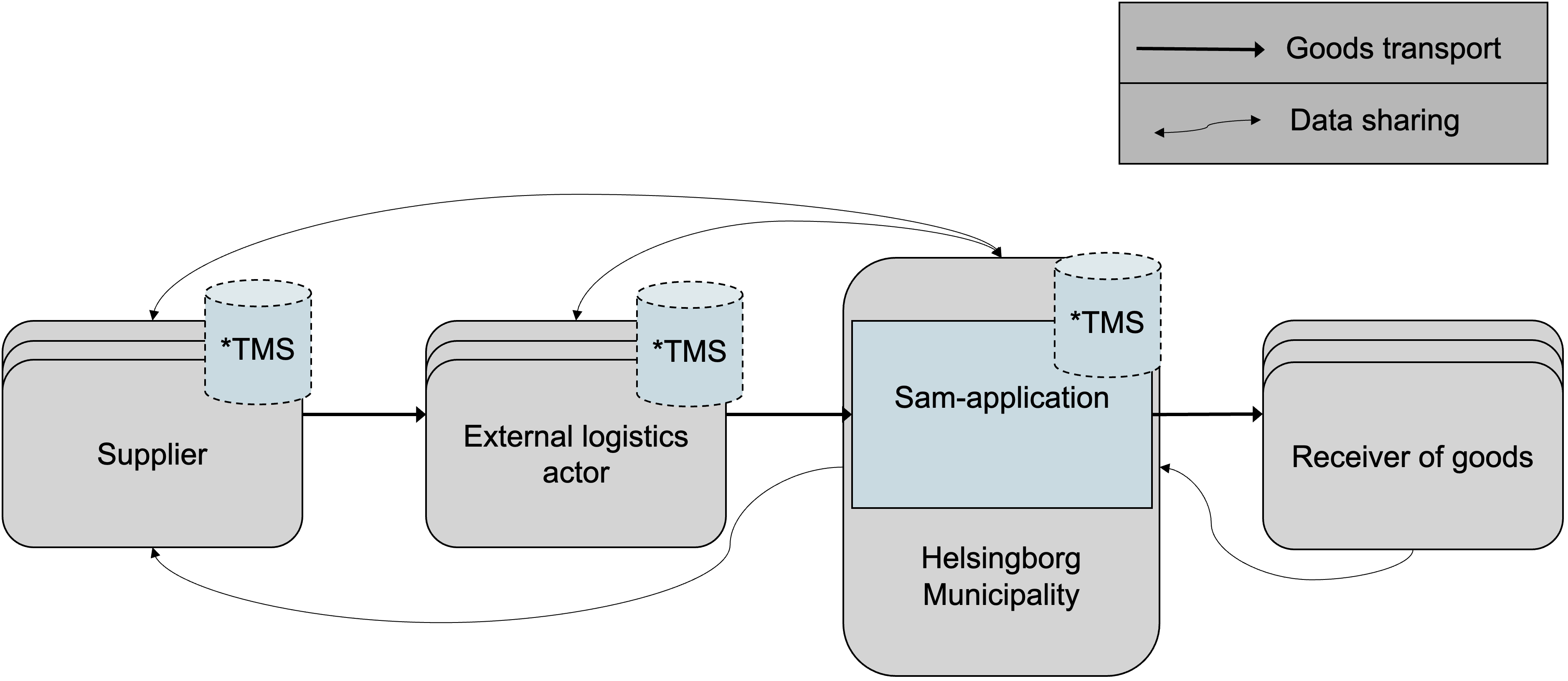}
    \caption{\revtwo{Actor model} of Helsingborg Municipality, illustrating different actors, systems, and a platform. The Sam-application, acting as a platform, is orchestrated by Helsingborg Municipality, and enables data sharing, supporting more efficient freight transport between involved actors.}
    \label{SAM-application}
\end{figure}

The Sam-application is illustrated in the case-specific \revtwo{actor model} (Figure~\ref{SAM-application}), where some involved actors, such as external logistics actors and suppliers, defined in Table~\ref{tab:definition-of-conceptual actors}, utilize the Sam-application to share and access logistics data. The sharing of data is enabled through registration of vehicle movements in different zones, thus facilitating the collection of real-time delivery data. This data allows for more efficient last-mile transportation by co-loading between logistics actors and informed decisions when planning the transports. \revtwo{The actor model shows that} a TMS can be connected to involved stakeholders, supporting logistics operations. 

As the municipality has an annual procurement volume of approximately SEK 4 billion for goods and services, the municipality operates within a context where transport efficiency plays a key role in supporting efficient and accountable public spending.

\changed{\subsubsection{Södertörn Municipalities}}

\begin{figure} [htbp]
    \centering
    \includegraphics[width=0.75\linewidth]{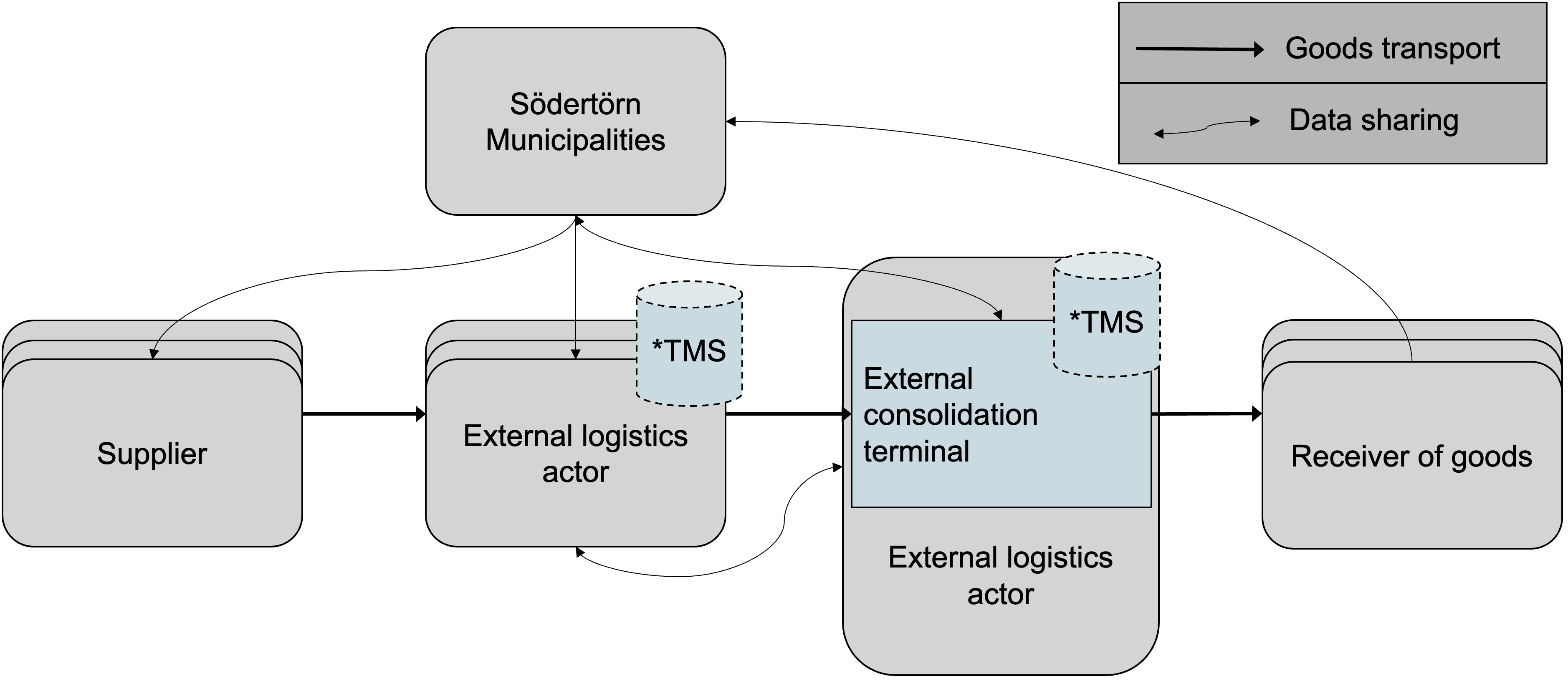}
    \caption{\revtwo{Actor model} of involved actors, platform, goods, and data sharing involved in the case of Södertörn Municipalities. The external consolidation terminal, acting as a platform, is orchestrated by an external logistics actor.}
    \label{Södertörn Municipalities}
\end{figure}

The Södertörn Municipalities, comprising eight local municipalities in the Stockholm region, utilize an external consolidation terminal to manage their contracted municipal transport operations. As illustrated in the case-specific \revtwo{actor model} (Figure \ref{Södertörn Municipalities}), this consolidation terminal serves as a central component in the logistics flow, acting as a platform where goods from various suppliers are received and redistributed to actors such as schools and public institutions.
The external consolidation terminal, supported by a connected TMS, enables vehicles to be optimized in their fill rate, thus contributing to the efficiency of municipal last-mile logistics. 

\begin{figure} [!htbp]
    \centering
    \includegraphics[width=0.75\linewidth]{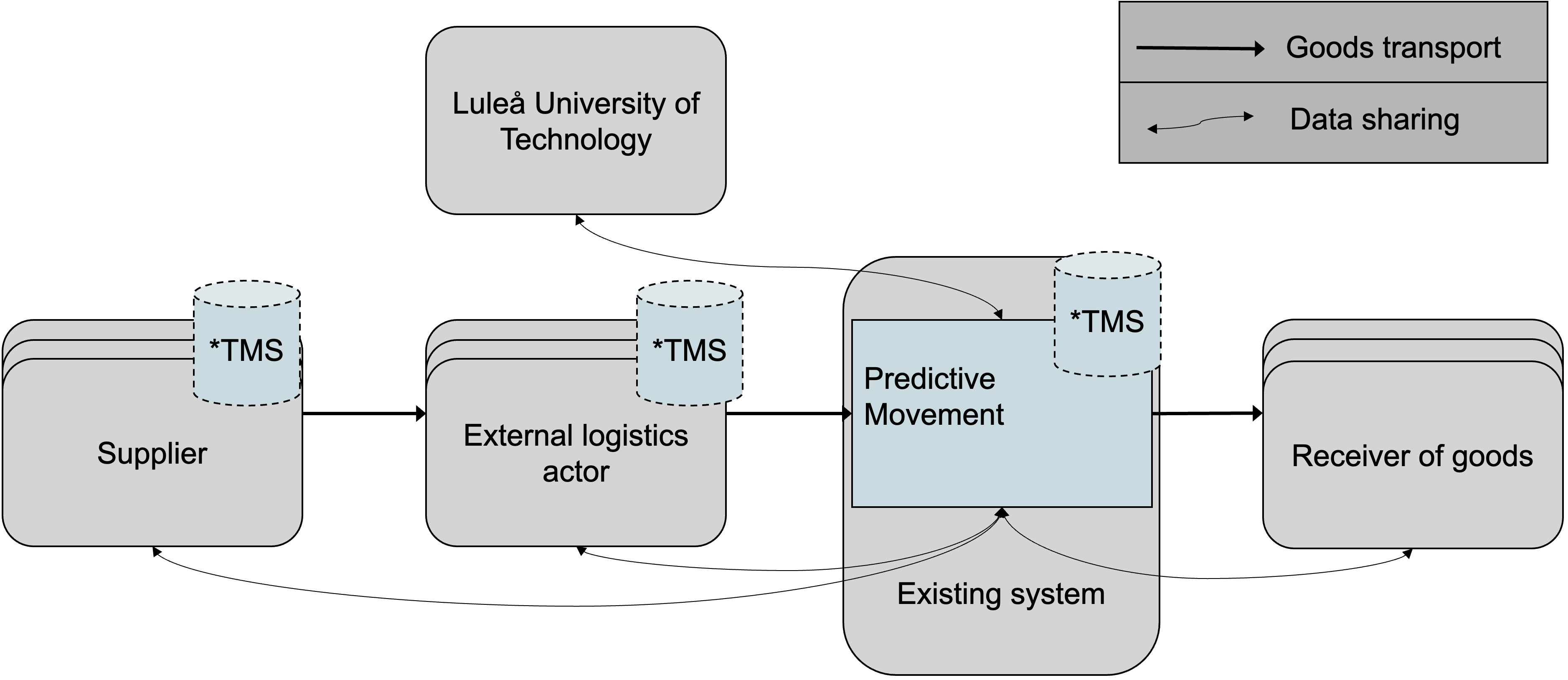}
    \caption{\revtwo{Actor model} of the initiative Predictive Movement supported by Luleå University of Technology, illustrating different actors, systems, and a platform. Predictive Movement acts as a platform and enables data sharing between involved actors, supporting more efficient logistics operations.}
    \label{Predictive Movement}
\end{figure}

\changed{\subsubsection{Predictive Movement}}
%Luleå University of Technology is contributing to the development of the initiative 
Predictive Movement is an open-source platform aimed at improving logistics efficiency through data sharing. Compared to the Sam-application, the platform can be considered less mature since it is still under development. However, it has been tested in several pilot projects, with actors such as transport operators and municipalities. The platform primarily focuses on rural areas and explores how logistics operations can be optimized in these contexts. The platform supports three main application areas: freight transport, public transit, and on-demand mobility services. As illustrated in the case-specific \revtwo{actor model} of Predictive Movement (Figure \ref{Predictive Movement}), the platform enables stakeholders, such as external logistics actors or suppliers, to share information on availability and demand for freight transport. This collaborative exchange of data results in a more efficient flow of goods. Furthermore, the platform supports dynamic coordination between actors to improve service levels without increasing the number of transport operations. The platform is also designed to be interoperable with existing systems, such as TMS, and can be adapted or extended by stakeholders with sufficient technical capabilities.

\section{Results}
\label{sec:results}

\revtwo{The results from the analysis of the empirical data are threefold. First, we synthesize the three actor models into a general one, embodying the variations observed (Section~\ref{sec:Actor model}).}
\revtwo{Next, we identify patterns in how data sharing is organized and enabled within ODE\textsubscript{PPF}. These patterns are derived from a deductive thematic analysis, resulting in a conceptual model% in Figure~\ref{fig:Analysis model}
, which illustrates how themes and sub-themes are connected to the aspects in the findings (Section \ref{sec:ConceptualModel}).} 
\revtwo{Thirdly, we sketch propositions for ODE\textsubscript{PPF} as a first step towards theories about their functioning (Section~\ref{sec:Propositions}).}

\subsection{Actor model}
\label{sec:Actor model}

\begin{figure} [tbp]
    \centering
    \includegraphics[width=0.85\linewidth]{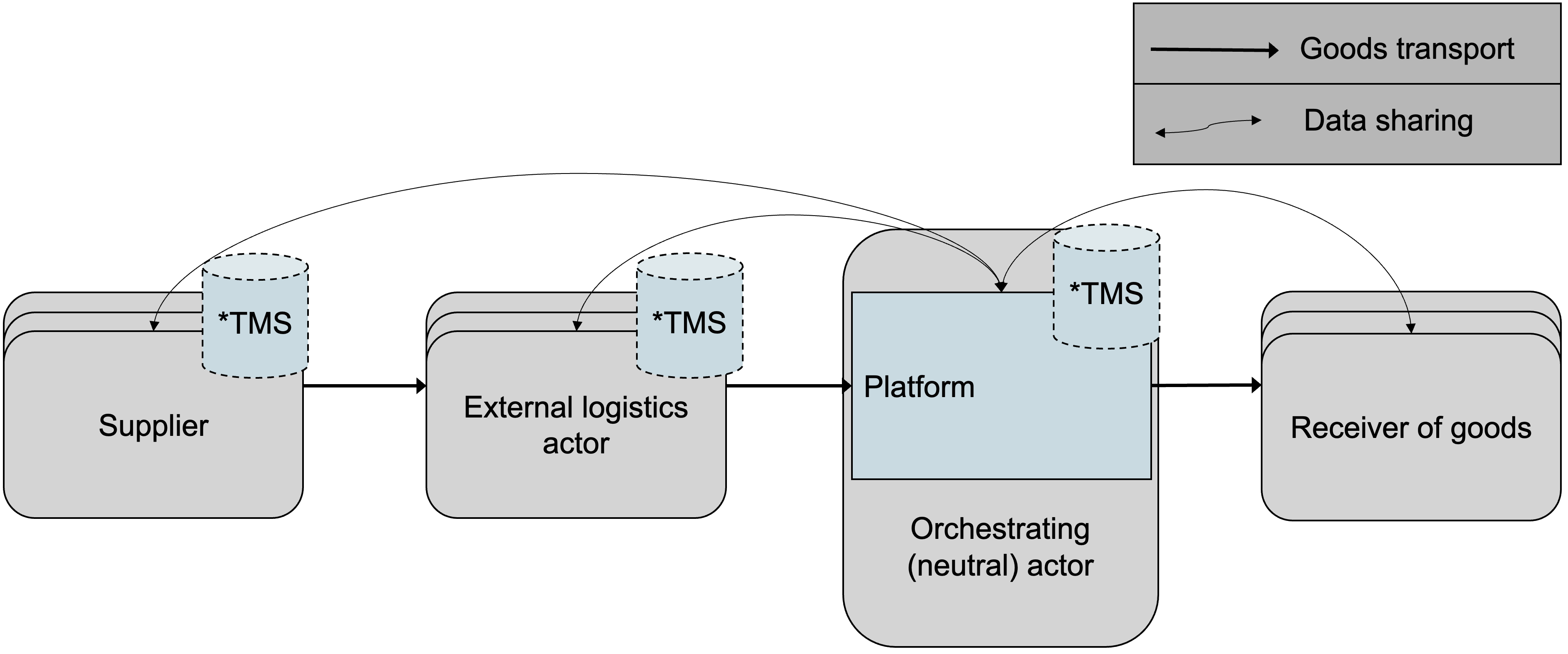}
    \caption{General \revtwo{actor model} of data sharing and exchange of goods between actors \changed{from a holistic perspective}. The thinner arrows represent data sharing between involved actors. The darker lines between the supplier, the external logistics actor, and the receiver of goods illustrate the flow of goods between the actors. The existing system and transport TMS can be seen as a separate system that supports the execution of logistical services.}
    \label{fig:conceptual}
\end{figure}

The general \revtwo{actor model} in Figure~\ref{fig:conceptual} visualizes a central data sharing platform and a set of actors and roles\revtwo{, as a synthesis of the case-specific actor models in Figures~\ref{SAM-application}--\ref{Predictive Movement}}. This includes an orchestrating (neutral) actor, supplier, external logistics actor, and receiver of goods. \changed{The model captures} the relationships between these actors, highlighting the exchange of both data and physical goods within the ecosystem. 

\changed{The model highlights} how data sharing is \changed{enabled through a combination of platform coordination and system integration. For instance,} data may be shared from the supplier or an external logistics actor to the platform. Through the platform, the orchestrating (neutral) actor and the recipient of the goods can access relevant data. 

\changed{A TMS may either be} an integrated component of the data sharing system or a parallel system that supports logistics interoperability. \changed{Based on the model, data sharing within ODE is not dependent on a single system or actor. Instead, it emerges through the orchestration between multiple actors, involving both technical and organizational components, and can therefore be understood as a socio-technical system.}

\revtwo{\subsection{Conceptual findings}}
\label{sec:ConceptualModel}

\changed{The findings are structured around the themes presented in  Table~\ref{tab:themes}, which} show recurring activities and patterns \revtwo{in the ODE\textsubscript{PPF}} related to how public and private actors interact with data sharing in the context of freight transport. 
\changed{The table also illustrates how each theme relates to the research questions. }

\begin{table}[t]
\centering
\caption{Recurring themes identified in the interviews, representing the codebook used in the thematic analysis, along with research questions related to each theme.}
\begin{tabular}{@{}p{3.1cm}p{8.5cm}p{2.4cm}@{}}
\toprule
\textbf{Theme} & \textbf{Description} & \textbf{Research question} \\
\midrule
1. Data sharing &
Exchange of data between actors such as municipalities, suppliers, and carriers, covering data sharing models in the three cases, as well as technical and organizational challenges. & 
RQ1 \\
2. Platforms &
Digital platforms to facilitate and offer structured solutions for data sharing, with focus on existing solutions and operational challenges, together with proposed approaches and system interoperability. &
RQ1 \\
3. Roles & Responsibilities and governance approaches of involved actors.  &
RQ2, RQ3 \\
4. Sensitive data & Aspects and challenges related to transparency, confidentiality, and legal compliance with data protection regulations. &
RQ2 \\
5. Data quality & Quality aspects related to accuracy, completeness, consistency/compliance, and timeliness. & RQ2  \\
6. Procurement & How public procurement influences data sharing as an enabler and a constraint. &
RQ3 \\
7. Sustainability aspects & Regarding the use of data sharing to integrate sustainability aspects into municipal freight logistics. &
RQ2 \\
\bottomrule
\end{tabular}
\label{tab:themes}
\end{table}

\begin{sidewaysfigure}
    \centering
    \includegraphics[width=1\linewidth]{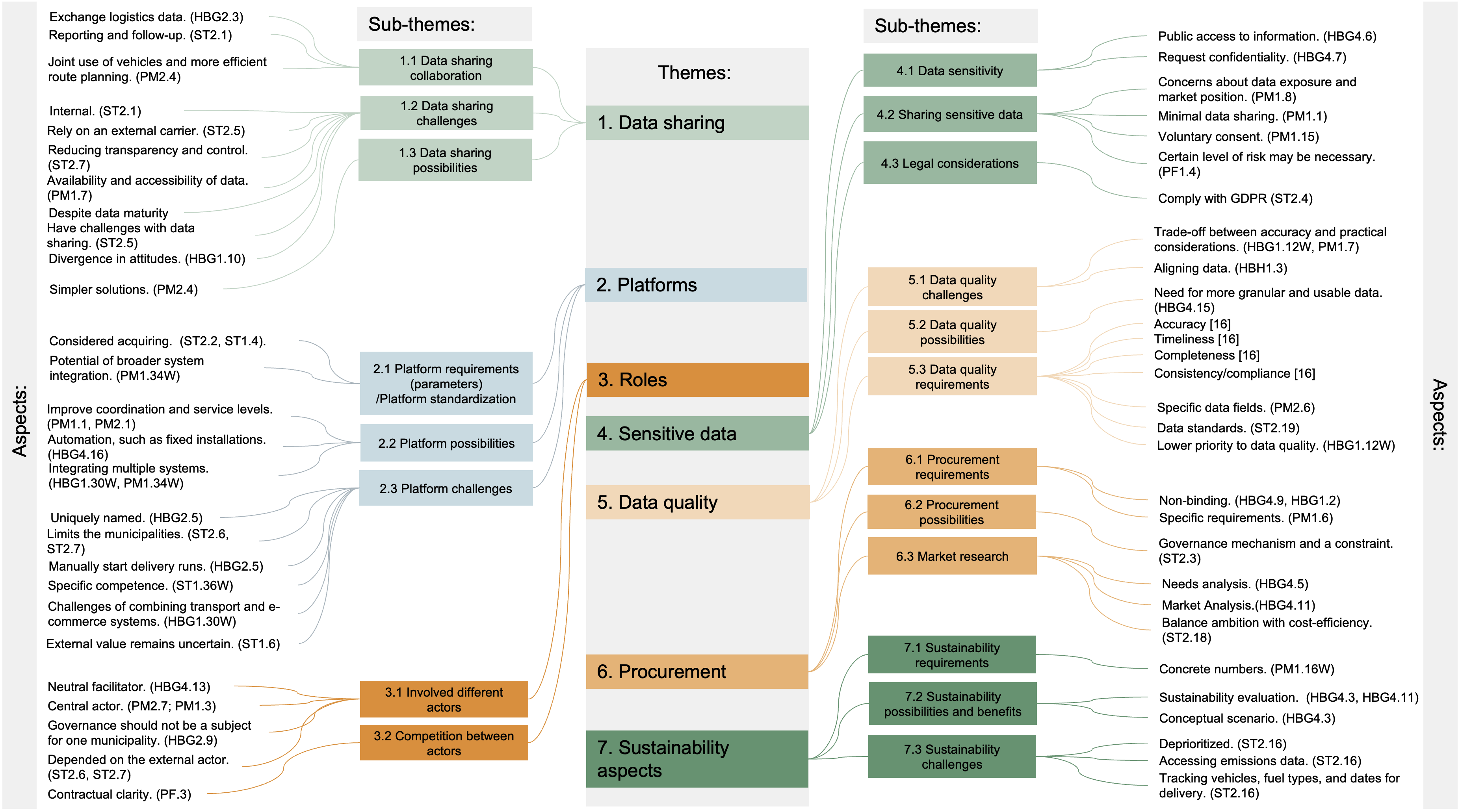}
    \caption{\revtwo{Conceptual model}, illustrating how themes \changed{(1-7)}, sub-themes \changed{(1.1-7.3)}, and analytical aspects are connected. The themes are located in the center of the figure, which is then followed by sub-themes. The themes are also defined in Table \ref{tab:themes}, illustrating the study’s code book. The analytical aspects are located based on the origin of each sub-theme, highlighting the connection to the findings of the study.}
    \label{fig:Analysis model}
\end{sidewaysfigure}

 In the following sections, each theme \changed{is elaborated, with reference to the analytical structure shown} in the \revtwo{conceptual model} (Figure~\ref{fig:Analysis model}). The \revtwo{conceptual model} illustrates the collected material from the interviews and how the different themes and sub-themes are connected and shape the dynamics of \revtwo{ODE\textsubscript{PPF}} in the analysis. \changed{References to information or quotes from individual interviewees are made using the ID's defined in Table~\ref{tab:respondents_participation}, to increase transparency of the results.}
\revtwo{Table~\ref{tab:ThemePerCase} summarizes the characteristic stressed for each case in the analysis, to guide the reader through the process.}

\begin{table}
\revtwo{
    \centering
        \caption{Characterization of key aspects for each theme per case in the analysis.}
    \begin{tabular}{{@{}p{3.1cm}p{3.1cm}p{3.1cm}p{3.1cm}@{}}}
\toprule
\textbf{Theme}	&	\textbf{Helsingborg}	&	\textbf{Södertörn}	&	\textbf{Predictive movement}	\\
\midrule
1. Data sharing	&	Integrated planning-based	&	Internal	&	Integrated market-based	\\
2. Platforms	&	Self-governed app	&	Externally operated	&	Self-governed platform	\\
3. Roles	&	Neutral facilitator	&	External actor	&	Central actor	\\
4. Sensitive data	&	Confidentiality concerns	&	GDPR compliance	&	Business concerns	\\
5. Data quality	&	High granularity	&	Completeness	&	Accuracy and timeliness	\\
6. Procurement	&	Feasibility	&	Cost	&	Enforcement	\\
7. Sustainability aspects	&	Integrated evaluation	&	Deprioritized implementation,	&	Monitoring	\\
\bottomrule
    \end{tabular}
    \label{tab:ThemePerCase}
    }
\end{table}

\vspace{5mm}

\subsubsection{Data sharing models}% in the three cases}

In the context of public sector freight logistics, \revtwo{\emph{data sharing} (1), as illustrated in the conceptual model} (Figure~\ref{fig:Analysis model}), refers to the exchange of transport-related data between actors, such as municipalities, suppliers, and carriers. This information may include vehicle positions sampled at certain events, goods data, and delivery updates. Data sharing \changed{functions as} a central mechanism in all three cases, \changed{but manifests in three distinct \revtwo{forms: \textit{integrated planning-based} data sharing in Helsingborg Municipality, \textit{internal} data sharing in Södertörn Municipalities, and \textit{integrated market-based} data sharing in Predictive Movement.}} \changed{These data sharing models reflect variations in the platform's maturity level and coordination among actors }\revtwo{while encapsulated in the actor model. We analyze three aspects of data sharing models: }\changed{\emph{data sharing collaboration} (1.1), \emph{data sharing challenges} (1.2), and \emph{data sharing possibilities} (1.3). Together, these sub-themes provide a structure for analyzing how the data sharing is configured in the three cases.}

%\revtwo{\subsubsection{Collaboration approaches}}
In Helsingborg Municipality, data sharing is facilitated by the municipality-owned data sharing platform, the Sam-application, which contrasts with Södertörn Municipalities, where data sharing is primarily \textit{internal}, \revtwo{although outsourced to a private actor}. The Sam-application enables actors to \textit{exchange logistics data} using geofencing technology (HBG2.3)\changed{, illustrating a more interactive \emph{data sharing collaboration} (1.1). In contrast, data sharing in Södertörn Municipalities} is \changed{primarily internal}, supporting the municipal organization with activities such as \textit{planning and reporting} (ST2.1). The system primarily functions as a procurement tool rather than a logistics platform for external actors. Although goods tracking is technically possible in Södertörn, the municipalities \textit{rely on an external carrier} to facilitate transport notifications (ST2.5). This dependency prevents real-time updates, \textit{reducing transparency and control} over transport data flows (ST2.7).

Predictive Movement represents \revtwo{a highly integrated model. Its} platform enables actors to share information on available transportation resources and delivery needs, facilitating the \textit{joint use of vehicles and more efficient route planning}. As a respondent explained: ``You have smaller actors who might not have their own individual vehicles, or they have their own individual vehicles, but they simply do not have enough goods to use them efficiently'' (PM2.4). By synchronizing their bookings, actors can jointly operate fleets, emphasizing the platform's role as a central data hub  \changed{and \emph{possibilities for data sharing} (1.3).} \changed{However, the model} depends on \textit{availability and accessibility of data}, \changed{as incomplete data limits coordination}: ``If the information about the transport is not complete, [...] then no one can coordinate it. It is like saying my house is for rent, but never saying which days'' (PM1.7). 

\revtwo{The actor model in Figure~\ref{fig:conceptual} captures the common elements and relations in all three cases. However, the practical implementations vary significantly, in level of integration, orchestrator roles, and consequently, benefits from the data sharing, as shown in Table~\ref{tab:ThemePerCase}.}

%\subsubsection{Technical and organizational challenges}
\emph{Data sharing} involves both technical and organizational \emph{challenges} \changed{(1.2)} in the three cases. To enable efficient and automated data sharing, the data format must follow established requirements (ST2.5). However, smaller operators may \textit{lack the technical skills} to interact with APIs. For them, \textit{simpler solutions} such as apps are preferable (PM2.4). \changed{At the same time}, larger actors, \textit{despite data maturity, have challenges with data sharing}: ``I would say that it very much depends on the data maturity of the supplier, which determines what they can deliver [...] it is not necessarily the case that the larger companies, which are data mature, are actually able to send the information either, because they are often very much locked into their systems'' (ST2.5). The interviews also revealed a \textit{divergence in attitudes} towards data sharing among logistics actors, while some stakeholders were positive, others saw additional data-sharing requirements as burdensome for resource-limited organizations: ``Many of these companies are already struggling today [...] when you bring up something new, it just becomes too much'' (HBG1.10). The technical and organizational challenges in the three cases are also illustrated in the general \revtwo{actor model} (Figure~\ref{fig:conceptual}), which visualizes how both \textit{data sharing} and the \textit{flow of goods} are structured. 

\begin{framed}\noindent\textbf{Summary -- data sharing:}
Data sharing emerged as a central mechanism in all three cases, taking different forms depending on technical maturity and organizational structure among the involved actors. Across all cases, respondents emphasized the importance of having available and accessible data, which otherwise might constrain the functioning of the ecosystem.
\end{framed}

\subsubsection{Logistics data sharing platforms and operational challenges} 
\changed{Logistics data sharing \emph{platforms} (2), as illustrated in the \revtwo{conceptual model} (Figure ~\ref{fig:Analysis model}), refer} to the design and implementation of digital platforms supporting municipal freight logistics. The design \changed{of these platforms varies} significantly between cases, revealing both \textit{technical opportunities and operational challenges}. \changed{These variations in platform design are reflected in how data sharing is implemented, influencing the integration and operation of data sharing in the three cases. These differences are closely related to the data sharing models identified in the previous section, influencing the orchestration of the data sharing.} 

In Helsingborg Municipality and Predictive Movement, the platforms are designed to \changed{support more integrated data sharing approaches among actors} such as carriers (external logistics actors), logistics operators (external logistics actors), and suppliers. In Helsingborg Municipality, the digital platform is integrated through the \emph{self-governed} Sam-application, which functions as a digital infrastructure for sharing transport data within the municipality and among external actors, \changed{indicating a more mature \emph{platform standardization} (2.1)}. The application also enables logistics operators to define their own distribution zones. As one respondent explained:  ``The carrier sets up their own zones [...] then the application automatically detects that they are probably picking up goods'' (HBG2.5). However, because zones are \textit{uniquely named} by each company, coordination becomes ``a bit of a guessing game'' (HBG2.5), \changed{highlighting \emph{platform challenges} (2.3).} 

In contrast, Södertörn Municipalities rely on an \revtwo{\textit{externally operated}} carrier \changed{acting} as a data intermediary. This setup \textit{limits the municipalities'} full visibility into historical data and delivery reports and prevents independent changes,  \changed{creating constraints (2.3)} by relying on the carrier (ST2.6; ST2.7). Predictive Movement \changed{represents a more collaborative, open-source platform-based approach, enabling coordination} of goods transport, public transport, and on-demand mobility. It aims to \textit{improve coordination and service levels} \changed{(2.2)} without increasing transport operations (PM1.1; PM2.1), and also functions as an intermediary layer that integrates with existing systems (PM1.3). Its open data architecture allows municipalities with technical capacity to adapt the system to specific requirements, while others may rely on a more general solution: ``if you have someone skilled in a municipality [...] then it is possible to do it'' (PM1.10). 

While digital platforms offer structured solutions for data sharing, their practical implementation reveals \textit{operational challenges} \changed{(2.3)}. In Helsingborg Municipality, for example, drivers must \textit{manually start delivery runs} in the app and ensure that the vehicle is within the designated zone, which is not straightforward: ``We do not know exactly where they stop [...] they might stop 50 meters away, so the zone has to be generously defined'' (HBG2.5). Given these operational challenges, several respondents (HBG4.16; PF1.9) stressed the importance of simplicity: ``Even the smallest manual task [...] becomes a risk that it will not be done'' (HBG4.16). Proposed, but not yet implemented solutions include \textit{automation, such as fixed installations} in vehicles (HBG4.16). The existing solutions and operational challenges related to logistics data sharing platforms are also illustrated in the general \revtwo{actor model} (Figure~\ref{fig:conceptual}). For example, \revtwo{limited data sharing between the platform and the receiver of goods may reduce visibility into delivery reports.} 

\changed{To address the limitations of existing platform implementations, the findings highlight the importance of interoperability and more integrated system architectures.} Given these limitations, Södertörn Municipalities has \textit{considered acquiring} its own TMS or delivery platform (ST2.2, ST1.4). A platform could also be procured together with other municipalities, since acquiring a platform requires \textit{specific competence}, which may not always be available locally (ST1.36W). Such a \textit{solution} could help address the lack of standardization underlying the operational challenges associated with existing platforms. It might include a dedicated Electronic Data Interchange (EDI) hub shared across municipalities to standardize communication and reduce costs (ST2.3). However, its \textit{external value remains uncertain}: ``There are solutions for this. But in the end, it is a matter of cost. Do we get the benefits back? Do the benefits outweigh the costs?'' (ST1.6).

In addition to acquiring their own platform, respondents emphasized the importance of \textit{integrating multiple systems}, which is illustrated in the general \revtwo{actor model} (Figure \ref{fig:conceptual}), to improve interoperability. However, one respondent reflected on the \textit{challenges of combining transport and e-commerce systems}: ``We are about to procure an e-commerce system [...] and I do not know how to handle transport in these systems. [...] If that’s not possible, then maybe we need to have something parallel [...] transport separately and the e-commerce system separately. [...] somehow integrate them with each other'' (HBG1.30W). Another respondent expanded on this idea, emphasizing the \textit{potential of broader system integration}: ``It is not just the TMS that is integrated with this platform, but several of these systems [...] It would be great if the transport order were created directly from the purchasing process and it is integrated'' (PM1.34W). The integration of the TMS is also illustrated in the general \revtwo{actor model} (Figure \ref{fig:conceptual}) through a cylinder-shaped representation. This figure illustrates that the TMS can either be integrated with other systems, such as the platform, or run in parallel.

\begin{framed}\noindent\textbf{Summary -- platform:}
Across the cases, respondents emphasized the importance of interoperable systems to support more efficient logistical coordination. The design and functionality of digital platforms for municipal freight logistics vary across the cases, revealing both technical opportunities and operational challenges.
\end{framed}

\subsubsection{Roles and Governance Approaches}

\begin{figure}[htbp]
  \includegraphics[width=1\linewidth]{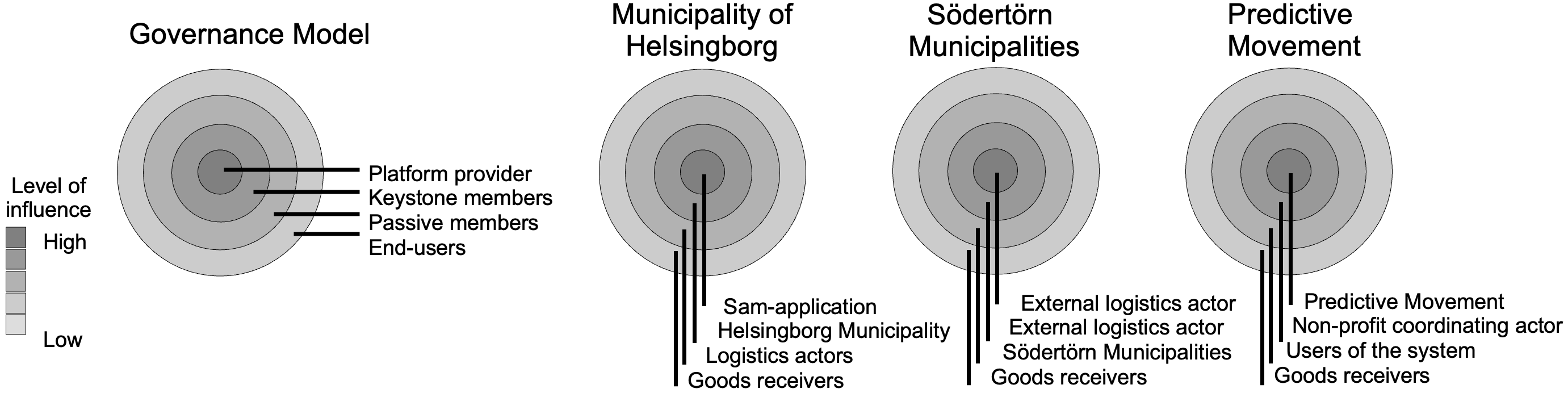}
    \caption{Overview of various governance models across the three cases, where each is placed in the governance model inspired by the ``onion model'' by Nakakoji et al.~\cite{Nakakojietal2002} and according to Runeson et al.~\cite{Runesonetal2021}. Helsingborg Municipality (Sam-application) and Predictive Movement are placed as central actors, whereas in the case of Södertörn Municipalities, the external actor acts as the central role, which limits the control of the municipalities on the platform. Other ecosystem actors, such as logistic actors, municipalities, users of the systems, and receivers of goods, are also placed in different layers, depending on their influence on the ecosystem.}
    \centering
    
    \label{fig:governance-models}
    
\end{figure}

\changed{Building on prior research (Section \ref{sec:Governance structure of ODE.}), we find that governance structures and roles differ across the cases within the ecosystems. These differences shape how control, coordination, and participation are organized}, reflecting \textit{different responsibilities and governance approaches.} More specifically, we identify three distinct governance roles: \textit{neutral facilitator}, \textit{external actor} and \textit{central actor}, including \emph{involvement of different actors} \changed{(3.1).} These influenced how collaboration, control, and \emph{competition between} the involved \emph{actors} \changed{(3.2)} are distributed among the actors in the ecosystem.

In the Helsingborg case, the municipality acts as a \textit{neutral facilitator} using the Sam-application to foster participation: ``The idea with this has been that the companies themselves should see an opportunity [...] we as a municipality, and perhaps as a more neutral party, can serve as a good initial step'' (HBG4.13). The neutral position helped reduce competition and foster engagement among involved actors: ``It becomes less competition-focused from the outset […] compared to if it had been profit-driven organizations'' (HBG4.14). The governance approach of Helsingborg Municipality using the Sam-application is also reflected in the governance model (Figure \ref{fig:governance-models}). In this model, the Sam-application and the municipality serve as a central point and keystone actor with strong influence on the ecosystem, fostering engagement for other actors involved in the ecosystem. However, these other actors have less influence on the ecosystem than the Sam-application and Helsingborg Municipality, as they are further away from the central governance.  

In contrast to the neutral facilitation in Helsingborg Municipality, Predictive Movement emphasized the need for a \textit{central actor} to manage the system, ideally a non-profit entity such as a university, a governmental institution, or a municipality (PM2.7; PM1.3). This governance structure of Predictive Movement is also illustrated in the governance model (Figure \ref{fig:governance-models}), where Predictive Movement has a central role, and a non-profit organization, such as a university or a municipality, acts as a keystone member, managing control over the ecosystem. However, others argued that the \textit{governance should not be a subject for one municipality}: ``It is not about a small municipality driving this, but rather that it should be a national or even EU-level requirement [...]'' (HBG2.9). This highlights expectations for interoperability and standardization to be driven by centralized actors, rather than municipalities alone. 

Another governance approach was evident in Södertörn, where the municipalities rely on the \emph{external actor} to manage freight logistics operations. Data sharing is used for internal reporting and follow-up purposes (ST2.1), and the municipality \textit{depends on the external actor} for different logistical services (ST2.6; ST2.7). This, in turn, made the municipality contractually bound to the external actor for logistical services. This governance structure is also reflected in the governance model (Figure ~\ref{fig:governance-models}), where the external actor assumes the central role (first layer), thus restricting the control (second layer) of municipalities in the ecosystem. The municipalities are therefore placed in the third layer due to the limitations imposed in the second layer. For an external perspective, a private actor (PF1.3), emphasized the role of \textit{contractual clarity} \changed{(3.2)} to have secure and transparent collaboration. From their perspective: ``In these municipal agreements [...] it is very clear what will happen going forward [...] without a very clear agreement, the supplier might have tried to look there [...] and pick the raisins out of the cake'' (PF1.3).

\begin{framed}\noindent\textbf{Summary -- roles and governance approaches:}

Roles concerning data sharing vary across cases, reflecting different responsibilities and governance approaches. The cases acted as neutral facilitators, central actors, or contractual dependents. These approaches shape how collaboration and control are distributed among actors in the ecosystems, influencing the coordination of data sharing and logistics operations. 
\end{framed}

\subsubsection{Sensitive data}
\changed{Handling of sensitive data introduces tensions between \textit{transparency, confidentiality, and compliance with data protection regulations such as the GDPR} \cite{EU_GDPR_2016}, shaping how data sharing can be implemented in the cases. These tensions are reflected across the cases, where the handling of \emph{sensitive data} (4.1) and \emph{legal considerations} (4.3)} emerged as recurring concerns. Municipalities operate under the principles of \textit{public access to information}, which can create tension with private actors' need for secrecy. As one municipal representative from Helsingborg Municipality explained, \revtwo{these \textit{confidentiality
concerns} are related to public access to information}: ``contract with a municipality can be requested as public information [...] it is already quite transparent'' (HBG4.6). While companies can \textit{request confidentiality} during public procurement processes, such requests are not always granted, as secrecy assessments are required (HBG4.7).

Similar concerns were raised by other respondents in Södertörn Municipalities, who emphasized removing certain types of data, such as receiver registers and names, for \textit{GDPR compliance} (ST2.4). \revtwo{Confidentiality also depends} on whether the data is commercially sensitive (ST2.10). Private actors can consider this data openness as a risk. A supplier involved in the Sam-application explained that municipal contracts can be difficult to interpret without direct involvement: ``If you are not actively involved in the contract [...] you might read the documents and think, I would never have signed this, but that is not how it works in practice'' (PF1.3).

\textit{Business concerns} regarding \textit{data exposure and market position} were also raised in the Predictive Movement case: ``It is this uncertainty about what it means for my own business […] they do not want to lose their place in the market'' (PM1.8). To mitigate these risks, actors in the cases described strategies to \emph{share sensitive data} \changed{(4.2).} One strategy emphasized was \textit{minimal data sharing}, where only the information strictly necessary to complete a task should be available (PM1.1). \textit{Voluntary consent} was also highlighted as a key mechanism, allowing users to approve of data sharing under specific conditions:  ``I think this idea of voluntarily sharing your data […] is something we should not forget'' (PM1.15). Despite these concerns, some respondents acknowledged that a \textit{certain level of risk may be necessary} to enable progress. As one supplier explained: ``Of course, I expose myself as a supplier, and that is the risk I have to take in exchange for a potential gain'' (PF1.4).

\begin{framed}\noindent\textbf{Summary -- sensitive data: }
The handling of sensitive data emerged as a concern related to transparency, confidentiality, and compliance with regulations. 
Municipalities operate under the principle of public access, which can create tension with private actors' need for secrecy. Strategies such as minimal data sharing and voluntary consent were emphasized to mitigate challenges and foster collaboration. 
\end{framed}

\subsubsection{Data quality}
\changed{Data quality} \revtwo{is a joint concern} \changed{across the cases and functions as a key component for effective data sharing, shaping how data can be used for coordination and decision-making within the ecosystem.} \revtwo{ Still, there are variations which w}\changed{e analyze, using four key dimensions of data quality \cite{Strongetal1997, ATTARD2015399}:} \textit{accuracy, completeness, consistency/compliance, and timeliness}. These dimensions are therefore used as an analytical framework for the different quality variations among the cases. 

In terms of \textit{accuracy}, a respondent in Helsingborg municipality emphasized \revtwo{\textit{high granularity}, especially the} \textit{need for more granular and usable data}\changed{, highlighting \emph{data quality possibilities} (5.2). However,} the Sam-application is considered a valuable initial step, \changed{but} lacks data: ``I would have liked to get more information related to fuel type, the company, and the individual employee driving'' (HBG4.15). This illustrates how data quality supports technical integration, environmental, and social goals. However, another respondent from Helsingborg municipality gave \textit{lower priority to data quality}: ``Data quality is the aspect we have focused on the least. [...] We want to get this started, to begin collecting and sharing data. [...] We may have geofencing zones [...] that are not entirely accurate [...] but that is not what matters most to us at this stage'' (HBG1.12W). \changed{This highlights a trade-off (5.1) between data quality and the volume of data, where data with missing details may be acceptable to enable collaboration.}

Similar challenges in managing the \textit{trade-off between accuracy and practical considerations} were evident in the case of Predictive Movement, where data quality is the foundation for their digital twins for transport simulations. However, limited real-world data often results in the use of synthetic data, raising questions about how accurate the data is: ``It is a question of how much you’re willing to accept'' (PM1.7). This concern relates to both \textit{accuracy and timeliness}, as outdated data can affect the trustworthiness of digital twins. 

Regarding \textit{completeness}, in Södertörn Municipalities, data quality depends on the external consolidation terminal and the level of data maturity among suppliers. Some actors provided detailed information, while others could only deliver limited input, such as Excel files or global location number (GLN), requiring enrichment by the carrier using internal systems: ``There are some actors who are very good [...] while others can [...] only send an Excel file with weight and a GLN number'' (ST2.5). Another respondent also emphasized the importance of \changed{\emph{data quality requirements} (5.3)} such as \textit{specific data fields}, \revtwo{including} origin, destination, pickup and delivery times, and goods dimensions, as inaccurate or missing data can cause operational problems (PM2.6).

For \textit{consistency/compliance}, to address the challenges with variation in maturity among suppliers, Södertörn Municipalities has adopted \textit{data standards}, developed with the Swedish Association of Local Authorities and Regions (SALAR)\footnote{\url{https://extra.skr.se/englishpages.411.html}}, although challenges remain (ST2.19). Issues related to standards were also noted in Helsingborg municipality, where \textit{aligning data} from multiple systems remains difficult. To facilitate integration, the municipality is exploring using GLN numbers: ``We have started looking into GLN numbers [...] so that we can combine and understand the data we receive'' (HBG1.3).

\begin{framed}\noindent\textbf{Summary -- data quality: }
Data accuracy, completeness, consistency/compliance, and timeliness \revtwo{are quality factors brought up in} the cases. 
These aspects can influence how the ecosystem functions in terms of efficiency and integration. In the studied cases,
actors balance ideal data quality against practical limitations and a willingness to share data.
\end{framed}

\subsubsection{Procurement}
\changed{Public procurement functions as both an enabler and a constraint for data sharing, shaping how requirements and contractual agreements influence} integration in municipal freight logistics. Across all cases, procurement influenced data sharing through \textit{specific requirements.} Interviews highlighted several key aspects, including \textit{needs analysis, market analysis, procurement possibilities, and the time-bound aspect of procurement agreements} (ST2.12). These key aspects are related to the Swedish Public Procurement Act (2016:1145), which regulates how public procurement should be conducted and addresses aspects such as transparency, competition, environmental, and social considerations. The \textit{needs analysis} was described as an initial step in the procurement process \changed{contributing to procurement-related \emph{market analysis}} \changed{(6.3)}, where the respondents emphasized the importance of questioning the need: ``Do we really need this, or are we just buying something we have always bought?'' (HBG4.5).

This \changed{market research} approach was then followed by the \textit{market analysis}, which aims to ensure \revtwo{the \textit{feasibility} of the requirements by determining whether they} are realistic and aligned with market capacity. A respondent explained: ``You ask questions to find out what is reasonable [...] we do not want to end up in a situation where the requirements are so strict that we get no offers'' (HBG4.11). Specifications must \textit{balance ambition with cost-efficiency}, highlighting the \textit{cost} considerations: ``You can specify anything. [...] it must not become too cost-driving'' (ST2.18). After needs and market analysis, procurement requirements have a central role in managing the behavior of involved actors and influencing data sharing \changed{(6.1)}. \changed{As} these requirements are often \textit{non-binding} (HBG4.9, HBG1.2), \changed{they do not enforce compliance among stakeholders}: ``We have started setting requirements to collect data. [...] we had not made it a mandatory requirement, only a recommended one. [...]'' (HBG1.2). As these are non-binding, they do not force compliance, and this complexity was further reflected by one respondent, who explained that despite municipalities having direct control, procurement still proves to be a challenging process: ``Procurement might be considered in the area where we have the greatest control […] but I still find it very difficult, because it requires getting many stakeholders on board and changing working practices'' (HBG1.14W). \changed{This reduces the likelihood of data sharing among stakeholders, which leads to uneven participation across the ecosystem.} 

However, \textit{enforcement} through \textit{specific requirements}, such as providing logistical data, was seen as a solution: ``One could write that they must provide their driving data'' (PM1.6), while flexibility in data formats may improve compliance: ``We can not require them to use the app if they have another system that works just as well'' (HBG2.2). Feasibility also depends on the type of logistic actors, those who acquire their transport may face greater challenges (HBG1.2). Furthermore, procurement was described to function as both a central \textit{governance mechanism and a constraint}\changed{, highlighting procurement possibilities (6.2).} When municipalities are contractually bound to providers, it may limit their flexibility. However, to overcome this, a dedicated platform could allow Södertörn Municipalities to avoid dependency on a logistics service provider: ``For us, it would mean that we are not tied to one transport provider, but rather free'' (ST2.3). 

\begin{framed}\noindent\textbf{Summary -- procurement: }
Public procurement acts as both an enabler and a constraint for digital integration in municipal freight logistics. Procurement requirements have a central role in shaping the behavior of suppliers. Key aspects include needs analysis, market evaluation, and the time-bound aspect of procurement agreements.
\end{framed}

\subsubsection{Sustainability aspects}
\changed{Sustainability can be viewed from several perspectives, in which frameworks such as the European Sustainability Reporting Standards (ESRS) \cite{EuropeanCommission2023ESRS} address three categories: Environmental, Social, and Governance. In this study, however, the cases examined primarily focus on environmental aspects.} Data sharing \changed{enables} \textit{opportunities \changed{(7.2)} to integrate environmental sustainability \changed{objectives}, specifically those aspects related to reducing emissions and air pollution into municipal freight logistics}, \changed{but its implementation is limited by \textit{challenges} (7.3)} in relation to prioritization and data accessibility. 

Respondents described how data sharing can support more sustainable logistics practices. In Helsingborg Municipality, for example, there is an ambition \revtwo{for \textit{integrated evaluation} of sustainability aspects, including \textit{sustainability evaluation} in} procurement, planning, and follow-up (HBG4.3; HBG4.11). In Södertörn Municipalities, similar solutions were described, where sustainability monitoring could be enabled by \textit{tracking vehicles, fuel types, and dates for delivery.} However, \revtwo{the \textit{deprioritized implementation} of these efforts reflects how such efforts are \textit{deprioritized} (ST2.16) in practice}. Another challenge emphasized concerns \textit{accessing emissions data}, particularly when external carriers are involved: ``It would have been easier to get an overview [...] but it is a bit difficult. [...] If they send just that one pallet on a truck, then the environmental impact becomes very large'' (ST2.16).

While these challenges exist, one respondent described a \changed{\textit{conceptual scenario}} that involves calculating the sustainability impact of a specific delivery based on its volume, vehicle type, fuel consumption, and distance traveled, \changed{indicating \emph{sustainability requirements} (7.1), potentially enabling} more informed decisions when ordering transport services (HBG3.3). Emissions could also be reduced by adjusting delivery schedules (HBG4.3) or by improving vehicle fill rates (ST2.17). However, as another respondent explained, \revtwo{\textit{monitoring} and the communication of} \textit{concrete numbers} can sometimes be more effective than sharing abstract ones: ``It is almost more important to be able to show, okay, but we have reduction of vehicles by three, instead of being able to state the exact number of kilograms of CO2 saved'' (PM1.16W). 

\begin{framed}\noindent\textbf{Summary -- sustainability aspects:}
Data sharing can support more sustainable freight logistics practices by enabling integration and monitoring of key sustainability aspects at a detailed level. These aspects are, for instance, goods volume, vehicle type, fuel consumption, and distance traveled. 
\end{framed}

\subsection{Summary propositions}
\label{sec:Propositions}
\changed{As a step towards theorizing the phenomenon of public-private optimization of municipal and regional freight logistics, we derived the conceptual model of constructs and actors \revtwo{in ODE\textsubscript{PPF}}, as presented in Figure~\ref{fig:Analysis model}, \revtwo{and Figure ~\ref{fig:conceptual}. Here, we further summarize} the above-reported findings in terms of proposition sketches (PS1--PS7) \revtwo{based on the three examined cases}, to be refined and validated in future work, see Table~\ref{tab:proposition_sketches}.} \revtwo{Drawing further on the proposition sketches, Figure \ref{fig:propositionsODE} illustrates how our emerging findings can be related to the general concepts of ODE identified by Runeson et al. \cite{Runesonetal2021}, which are further elaborated below. }

\revtwo{
%TBD relation to ODE in Figure~\ref{fig:propositionsODE}
\begin{figure}[hbt]
    \centering
    \includegraphics[width=1\linewidth]{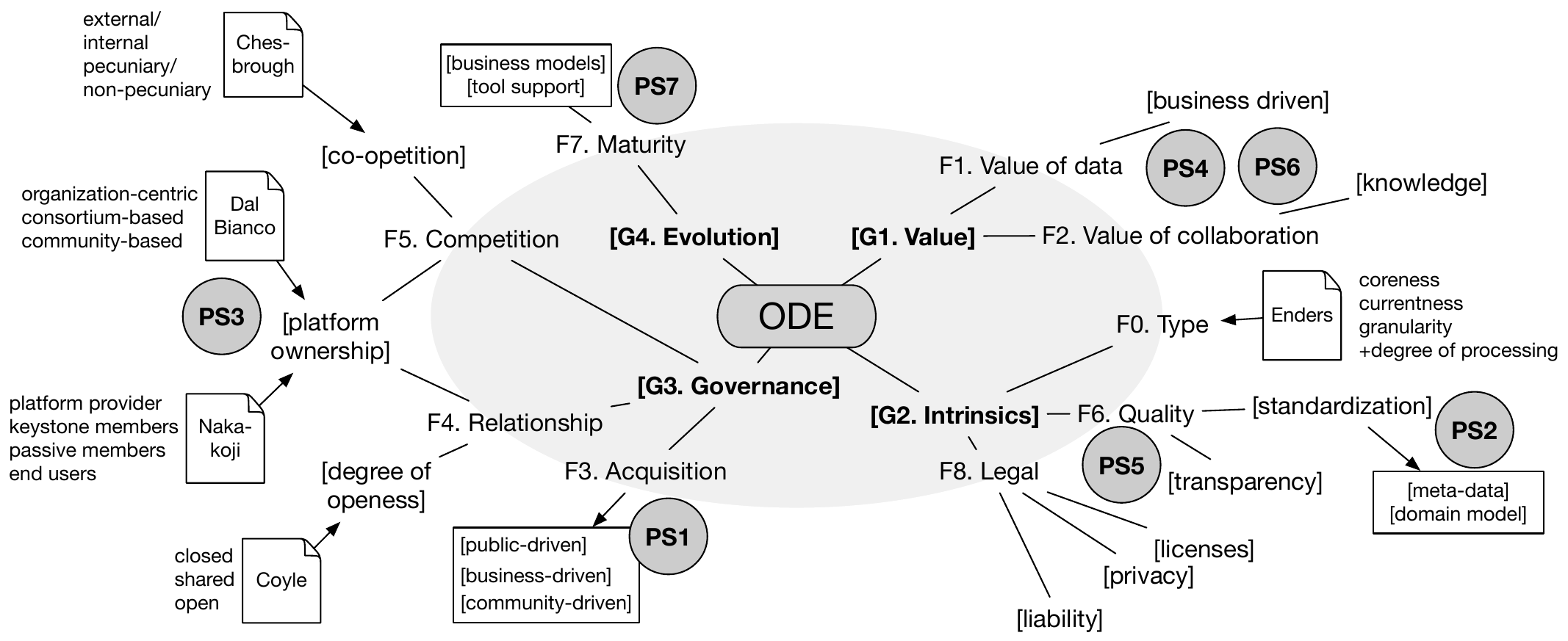}
    \caption{\revtwo{Proposition sketches from Table~\ref{tab:proposition_sketches} mapped to concepts of ODE \cite{Runesonetal2021}, illustrating the conceptual relations between the ODE\textsubscript{PPF} findings and our prior ODE research. The darker circles (PS1--PS7) represent the proposition sketches, G1--G4 represent the key conceptual 
    groups, while F0--F8 represent underpinning factors.}}
    \label{fig:propositionsODE}
\end{figure}
}

\begin{table}[htb]
    \centering 
    \begin{tabular}{cp{10cm}}
    \toprule
        \bf{ID} & \bf{Proposition Sketch}\\
         \midrule
        PS1 & Data availability and accessibility is at the core of freight logistics optimization.\\
        PS2 & Technical interoperability across systems and actors is foundational for data sharing.\\
        PS3 & Public actors may take different governance roles, but the central actor must be trusted as being fair.\\
        PS4 & Risks related to sensitive data must be balanced with benefits for actors.\\
        PS5 & Data quality impacts the data ecosystem efficiency and integration.\\
        PS6 & Public procurement requirements may enforce \emph{and} constrain data sharing.\\
        PS7 & Detailed transport data are needed to perform valid freight logistics optimization with respect to environmental sustainability.\\
         \bottomrule
    \end{tabular}
    \caption{\changed{Proposition sketches derived from the results above, and aimed for further refinement and validation in future work.}}
    \label{tab:proposition_sketches}
\end{table}

\hspace{5mm}

\changed{We synthesize across our surveyed cases that data availability and accessibility is at the core of the envisaged optimization of freight logistics (PS1).} \revtwo{This is an area of conflict between private and public actors in ODE\textsubscript{PPF}. The approach to data acquisition must combine public-driven and business-driven interests. }

\changed{The platforms play a key role in enabling \revtwo{sharing of freight logistics data}, where \revtwo{interoperability between platforms and existing systems, such as TMS,} is foundational (PS2).} \revtwo{Interoperability requires standardization, for example of data and meta-data formats and APIs.}
\changed{Platforms, however, only provide the technical means. Effective and trusted coordination and governance mechanisms are critical for facilitating and enabling any sharing and collaboration at large (PS3).} \revtwo{In ODE\textsubscript{PPF}, the public actor has the primary interest in initiating an organization-centric platform, which gradually may be community-based through open source software development, as in the JobTech case \cite{Runesonetal2021}. The public actor is a trusted actor, representing common good for the society.}

\changed{When sharing data, it is also necessary to balance sensitive data with} \revtwo{the value provided for actors (PS4), such as market exposure and efficiency improvements. Logistics data may be sensitive as, for example, fill rates of transports is a business health indicator. Risk mitigation for sensitive data includes procurement contracting, only sharing minimal amounts of data for the purpose, and protecting data by technical design.} 

\revtwo{Data quality affects the trustworthiness of the data ecosystem (PS5), as discussed in Section~\ref{sec:DataQuality}. Both public and private actors rely on data for their operations and business decisions. Transparency may contribute to trustworthiness, but must be balanced with respect to sensitive data, as indicated in PS4.} 

\changed{Public procurement requirements can enforce and constrain} \revtwo{ freight data sharing (PS6). However, if the requirements are too extensive, logistics actors may not bid on the call.} \changed{To reach freight logistics optimization, it is, however, necessary to have detailed transport data, influencing the environmental sustainability (PS7).} \revtwo{Ultimately, this is to be aligned with the business models of the logistics actors.}

\section{Discussion}
\label{sec:discussion}

In the following section, we discuss the findings in relation to prior research. The discussion is organized around three areas: platform configuration and design, enabling factors for data sharing, and aspects related to sustainable freight transport. 

\subsection{Platform configuration and design}
\changed{The findings show that platform configuration plays an important role in shaping how data sharing is structured and implemented in \revtwo{ODE\textsubscript{PPF}}. These findings address RQ1 by showing how relationships between involved actors in the data-sharing ecosystem are enabled through different platform configurations.} For regions and municipalities to achieve more efficient and sustainable last-mile freight logistics, data sharing is essential. It functions as a central mechanism that fosters integration \revtwo{among involved actors} and \changed{drives value creation, including operational value, such as improved routing and co-loading, as well as value in the form of increased transparency and coordination among actors.} 

\revtwo{These forms of value are also indicated in the examined cases. For example, Helsingborg Municipality and Predictive Movement illustrate how data sharing among actors can support coordination through activities such as} co-loading and more efficient route planning. \revtwo{On the other hand, Södertörn Municipalities illustrate how dependence on an external logistics actor can limit access to transport-related data, reducing the transparency among actors within the ecosystem. These observations} are also in line with Runeson et al.~\cite{Runesonetal2021}, who emphasize data value and collaboration around data as main business drivers for \changed{ODE. While Möller et al.~\cite{Moller2020} identify visibility and optimization as dominant \revtwo{data-driven services that form the basis for business models in} logistics, our findings \revtwo{provide a perspective on how such value is created in ODEs\textsubscript{PPF} in the explored domain of public-private municipal and regional last-mile freight logistics. We also emphasize} that, in the studied domain, value creation is not only enabled through such data-driven services, but also shaped by platform configurations, governance structures, and varying levels of technical maturity among actors. \revtwo{In more detail, our three cases indicate that platform configurations, as illustrated by the Sam-application and Predictive Movement, influence how data sharing is enabled, while being shaped by how these platforms are orchestrated and controlled. Furthermore, varying levels of technical maturity among logistics actors using these platforms may also affect their ability to integrate with the platforms and share data within the ecosystem.} \revtwo{These observations also complement Bartolomucci et al.~\cite{Bartolomucci2026}, who conceptualize data ecosystems as socio-technical systems in which collaborative and data governance are interrelated dimensions, where interplay and integration influence the ability to derive and deliver value from data. Whereas their analysis focuses on closed data ecosystems, our findings show how socio-technical aspects are evident in ODE\textsubscript{PPF}, and more specifically in public-private municipal and regional last-mile freight logistics.}}

While data sharing is crucial to ODE, its effectiveness depends on a technological platform architecture that enables interoperability among actors and systems, creating an environment that allows for data sharing. \changed{This finding is in line with Ofe and de Reuver \cite{OfeDeReuver2025}, who highlight interoperability as a key enabler of data sharing in data platform ecosystems, where our findings also indicate that limitations in interoperability \revtwo{between TMS and other logistics systems} may constrain data sharing and value creation in } \revtwo{ODE\textsubscript{PPF}.}

This aligns with prior research that highlights system interoperability as a key element for effective logistics operations \cite{PAN_et_al_2021}. This includes, for example, integrating systems such as \revtwo{TMS} and e-commerce systems. Furthermore, operational challenges exist, such as manual processes, a lack of automation, inadequate technical skills, and system lock-in. This further underscores the need for more interoperable \revtwo{platforms}, where the \revtwo{ODE\textsubscript{PPF}} can act as a technical framework \cite{Larsson_Runeson_Linåker_2024}, promoting standardized solutions and reducing limitations such as system lock-in among stakeholders.

\subsection{Enabling factors for data sharing}
In addition to the platform configuration and design, \changed{our findings identify a set of interrelated enabling factors \revtwo{in the context of freight logistics}, including data quality, legal aspects, procurement, and governance structures, that shape how data sharing is realized in practice. These findings address RQ2 by indicating that data sharing in \revtwo{ODE\textsubscript{PPF}} is not only a technical challenge but is also influenced by socio-technical and organizational aspects. These enabling factors are also identified in inter-organizational data sharing, such as challenges related to data quality, legal uncertainty, and coordination among actors \cite{Jussen2024}. Our findings contribute to this research area by showing how these challenges manifest and interact in the context of \revtwo{ODE\textsubscript{PPF}}, shaping how data sharing is realized in practice.}  

Data quality is a critical factor for data sharing, though a trade-off exists between volume and data quality, increasing the complexity of reaching a balance. Inadequate data quality can hinder the ecosystem from reaching its full potential \cite{Larsson_Runeson_Linåker_2024} due to its importance in terms of value and usability to the ecosystem. However, the studied cases revealed different requirements. While some respondents highlighted the need for a large volume of data for platform functionality, others emphasized the need for more granular and usable data, such as specific data fields (PM2.6) for transportation (HBG4.15). Given that both aspects can have an essential role in the functioning of the ecosystem, there is a need to consider the requirements depending on the scenario. \changed{This aligns with prior research by Strong et al.~\cite{Strongetal1997}, who emphasize that data quality is context-dependent and defined by its fitness for use. Our findings further suggest that actors within \revtwo{ODE\textsubscript{PPF}} have different priorities of data properties, such as volume or data quality, thereby shaping how data sharing practices are implemented across the ecosystem.}

In addition to data quality, legal aspects such as GDPR \cite{EU_GDPR_2016} play an important role in the ODE, yet uncertainty around legal requirements exists \cite{Runesonetal2021}, \cite{Rajamäe_Soosaar_Nikiforova_2025}. To mitigate this and encourage data sharing, strategies such as minimal data sharing and voluntary \revtwo{participation} can therefore be a solution, complemented by clearer legal guidelines \cite{Dulong_de_Rosnay_Janssen_2014}, \cite{Magalhäs2016}. \changed{These findings therefore indicate that legal uncertainty can create barriers, where unclear regulations can reduce willingness among actors to participate in data sharing.} 

Together with legal aspects, procurement emerges as both a central governance mechanism and a constraint. Although municipalities feel that procurement is an area where they have considerable influence, requirements related to data sharing are often non-binding. We therefore argue for having more binding requirements in the procurement agreements, since procurement \changed{functions as a critical governance mechanism, where binding requirements can influence the extent to which data sharing is realized \changed{among actors.} If non-binding requirements are used, this can result in reduced participation among actors and limited effectiveness of the \revtwo{ODE\textsubscript{PPF}}.} 

Regarding governance, it has an essential role in data sharing \cite{Bartolomucci2026}, as it contributes to ecosystem orchestration, establishes standardization, builds trust, and fosters collaborations among actors \cite{Larsson_Runeson_Linåker_2024}. Municipalities can assume the role of a keystone actor, having a larger impact on the ecosystem \cite{Jansen_2014}, \cite{Harrison2012}. For example, if the municipality acts as a neutral actor, it can reduce competition and orchestrate collaboration among stakeholders (HBG4.14) in various governance structures, including community-based or keystone-centric structures \cite{Runesonetal2021}, \cite{OliveiraMarcelo2019}. However, opinions vary on whether governance should be managed by local municipalities or centralized to a higher level, such as the EU. \changed{These findings emphasize that governance not only orchestrates the ecosystem but also influences control, shaping the collaboration among involved stakeholders.} 

\subsection{Sustainable freight transport}

Sustainable freight transport depends on data sharing to enhance logistics efficiency and \changed{enable} co-loading. \changed{However, our findings indicate that such data sharing is often constrained by limited data accessibility, insufficient} prioritization, and \changed{challenges in integrating data} into key structures such as procurement. \changed{Prior research has highlighted the importance of data accuracy for achieving low-carbon operations \cite{Melville2017}.} However, \changed{our findings indicate that in \revtwo{ODE\textsubscript{PPF}}, the challenge is not solely related to data accuracy but also to the integration and prioritization of sustainability-related data between actors.}  

\changed{While some respondents described combinations of data such as} vehicle type, fuel type, and distance traveled \changed{as useful for enabling more informed decision-making, these examples were often conceptual and not fully implemented in practice. This indicates that the key challenge lies not in identifying relevant data, but in establishing the standards by which such data can be integrated into operational processes, collected, and shared among stakeholders. Achieving environmental goals, including national \cite{trafikverket2025} and global sustainability targets \cite{SDGS2023}, highlights the need for data-sharing among actors to be prioritized. Accordingly, this shows that} ODE can contribute to more efficient and sustainable freight transport, provided that data is accessible and integrated within governance and procurement structures.

\section{Conclusion}
\label{sec:conclusion}

This study examined how data sharing practices \changed{are shaped} in municipal and regional last-mile freight logistics in Sweden. Through a qualitative \changed{multiple-case study} of three cases, the findings \changed{show} that data sharing \changed{functions} as a \changed{central coordination mechanism, enabling integration and value creation across actors (RQ1).} 

\changed{The findings contribute to research and theoretical perspectives on ODEs by \revtwo{providing empirical insights into how ODE\textsubscript{PPF} function in municipal and regional last-mile freight logistics, indicating that ODE\textsubscript{PPF} are influenced not only by} technical configurations, \revtwo{such as technological platforms,} but also by several interrelated socio-technical aspects}. \revtwo{These include} enabling factors such as data quality, handling of sensitive data, procurement, and governance structures (RQ2). These aspects shaped the development of an \revtwo{actor model} illustrating \revtwo{actors and activities that contribute to the} functioning of \revtwo{ODE\textsubscript{PPF}} in last-mile freight logistics. \changed{From a practical perspective, the study contributes by highlighting the importance of \revtwo{clear} governance structures, interoperable platform design, \revtwo{and data-sharing requirements in municipal procurement practices.}}

\changed{The findings further show that municipalities} can assume the role of neutral keystone actors (RQ3), orchestrating the ecosystem \changed{by coordinating actors and facilitating data sharing, for instance through procurement practices. This also suggests that having clear governance structures and responsibilities is critical for enabling efficient data sharing in public-private domains. The study also indicates that \revtwo{ODE\textsubscript{PPF}} have the potential to contribute to more sustainable freight logistics \revtwo{through co-loading activities}, provided that the relevant logistics data is accessible and usable for operational planning.}  

However, as this is an \changed{exploratory study}, the focus on Swedish municipalities and the limited sample size may restrict transferability to other contexts. \changed{The findings should be interpreted in this context, and therefore, we do not make any claim of universal generalization.} Future research should examine \changed{additional cases and governance models, both in public} and private \changed{domains, and further explore} practices related to data sharing, \changed{including platform design,} data quality, legal frameworks, governance structures, interoperability, and business models.


\section{Acknowledgments}

The research is funded by the Swedish Transport Administration in the Triple F program (Fossil Free Freight), Project nr 2022.5.2.5. We thank the anonymous reviewer for feedback that helped sharpen our analysis and writing.

\appendix
\section{Interview questions}
\label{app1}

Below are the interview questions used during the first round of semi-structured interviews conducted with the three cases. The questions are organized according to the research questions from which they were derived.

\changed{\textbf{Research Question 1 (RQ1):}} What are the conceptual relations and data sharing practices between actors in the last-mile freight transport of goods on behalf of public actors?
 
Do you share data within freight transport, and if so, how do you share data?
What are the main objectives of sharing data? What value does the data have for the parties involved?

What are the main objectives of accessing the shared data? What type of data do you usually share or access? 
How do you work with sensitive data? 

\changed{\textbf{Research Question 2 (RQ2):}} How do actors address quality control, legal compliance, sustainability, governance, and other aspects that they consider important?

How do you think the quality of shared data affects your business?
Do you have any kind of system or standard to ensure that the data shared is of high quality? If so, how do you work with that system?

Do you have any system to protect the integrity of shared data? If so, how do you work with that system? 

When you work with data sharing, are there any legal requirements/policies that affect how you work? How do you think it works, and does it need to be changed? Are there any legal barriers that affect how you work with shared data? 

Are there any other aspects related to data sharing in freight transport that you consider important? 

\changed{\textbf{Research Question 3 (RQ3):}} What specific roles do public actors undertake?

What roles do you think municipal actors take on in data sharing? We often talk about data consumers, producers, or platform providers. Is this something you recognize?
How do you work with competition between different actors and data sharing?
What roles do you see other actors playing in data sharing related to freight transport? 
If there is one, which actor do you think takes on the role of platform provider/main responsible party, i.e., enabling data sharing? Why do you think this actor takes on this role?

%%       \cite{<label>} ==> [1]

%%


\bibliographystyle{elsarticle-num} 
\bibliography{mybibliography}

@article{Altendeitering2024,
  title = {A design theory for data quality tools in data ecosystems: Findings from three industry cases},
  volume = {153},
  ISSN = {0169-023X},
  DOI = {10.1016/j.datak.2024.102333},
  journal = {Data \& Knowledge Engineering},
  publisher = {Elsevier BV},
  author = {Altendeitering,  Marcel and Guggenberger,  Tobias Moritz and M\"{o}ller,  Frederik},
  year = {2024},
  month = sep,
  pages = {102333}
}

@article{ATTARD2015399,
title = {A systematic review of open government data initiatives},
journal = {Government Information Quarterly},
volume = {32},
number = {4},
pages = {399-418},
year = {2015},
issn = {0740-624X},
doi = {10.1016/j.giq.2015.07.006},
author = {Judie Attard and Fabrizio Orlandi and Simon Scerri and Sören Auer}
}

@article{Bonina2020,
	author = {Carla Bonina and Ben Eaton},
	doi = {j.giq.2020.101479},
	issn = {0740-624X},
	journal = {Government Information Quarterly},
	number = {3},
	pages = {101479},
	title = {Cultivating open government data platform ecosystems through governance: Lessons from Buenos Aires, Mexico City and Montevideo},
	volume = {37},
	year = {2020}}

@article{BraunClarke2006,
author = {Virginia Braun and Victoria Clarke},
title = {Using thematic analysis in psychology},
journal = {Qualitative Research in Psychology},
volume = {3},
number = {2},
pages = {77-101},
year = {2006},
publisher = {Routledge},
doi = {10.1191/1478088706qp063oa}
}

@article{Dawes2016,
	author = {Dawes, Sharon S and Vidiasova, Lyudmila and Parkhimovich, Olga},
	doi = {10.1016/j.giq.2016.01.003},
	journal = {Government Information Quarterly},
	number = {1},
	pages = {15--27},
	publisher = {Elsevier},
	title = {Planning and designing open government data programs: An ecosystem approach},
	volume = {33},
	year = {2016}}

@Article{Harrison2012,
AUTHOR = {Harrison, Teresa M. and Pardo, Theresa A. and Cook, Meghan},
TITLE = {Creating Open Government Ecosystems: A Research and Development Agenda},
JOURNAL = {Future Internet},
VOLUME = {4},
YEAR = {2012},
NUMBER = {4},
PAGES = {900--928},
ISSN = {1999-5903},
DOI = {10.3390/fi4040900}
}

@ARTICLE{Immonen2014,
  author={Immonen, Anne and Palviainen, Marko and Ovaska, Eila},
  journal={IEEE Access}, 
  title={Requirements of an Open Data Based Business Ecosystem}, 
  year={2014},
  volume={2},
  number={},
  pages={88-103},
  doi={10.1109/ACCESS.2014.2302872}}

@ARTICLE{Lakomaa2013,
  author={Lakomaa, Erik and Kallberg, Jan},
  journal={IEEE Access}, 
  title={Open Data as a Foundation for Innovation: The Enabling Effect of Free Public Sector Information for Entrepreneurs}, 
  year={2013},
  volume={1},
  number={},
  pages={558-563},
  doi={10.1109/ACCESS.2013.2279164}}

@inproceedings{Magalhäs2016,
author = {Magalh\~{a}es, Gustavo and Roseira, Catarina},
title = {Exploring the Barriers in the Commercial Use of Open Government Data},
year = {2016},
isbn = {9781450336406},
publisher = {ACM},
doi = {10.1145/2910019.2910078},
booktitle = {Proc. 9th Int. Conf. on Theory and Practice of Elec. Gov.},
pages = {211–214},
numpages = {4},
location = {Montevideo, Uruguay},
series = {ICEGOV '15-16}
}

@article{Moradi2022,
	title = {A {Comprehensive} {Method} for {Improving} the {Quality} of {Open} {Government} {Data} and {Increasing} {Citizens}’ {Willingness} to {Use} {Data} by {Analyzing} the {Complex} {System} of {Citizens} and {Organizations}},
	volume = {2022},
	issn = {1076-2787},
	doi = {10.1155/2022/5876035},
	journal = {Complexity},
	author = {Moradi, Mohammad and Mazoochi, Mojtaba and Ahmadi, Mohammad},
	editor = {Murari, Andrea},
	year = {2022},
	publisher ={Hindawi},
	pages = {5876035},
}

@inproceedings{Oliveira_Lóscio_2018,
  author = {Oliveira,  Marcelo Iury S. and Farias Lóscio,  Bernadette },
title = {What is a data ecosystem?},
year = {2018},
isbn = {9781450365260},
publisher = {ACM},
doi = {10.1145/3209281.3209335},
booktitle = {Proc. of the 19th Ann. Int. Conf. on Digital Government Research: Governance in the Data Age},
articleno = {74},
numpages = {9},
address = {Delft, The Netherlands}
}

@article{Runesonetal2021,
title = {Open Data Ecosystems — An empirical investigation into an emerging industry collaboration concept},
journal = {Journal of Systems and Software},
volume = {182},
pages = {111088},
year = {2021},
issn = {0164-1212},
doi = {10.1016/j.jss.2021.111088},
author = {Per Runeson and Thomas Olsson and Johan Linåker}
}

@article{Slibar2021,
author = {Barbara \v{S}libar and Dijana Ore\v{s}ki and Nina {Begi\v{c}evi\'{c} Redep}},
title ={Importance of the Open Data Assessment: An Insight Into the (Meta) Data Quality Dimensions},
journal = {SAGE Open},
volume = {11},
number = {2},
year = {2021},
doi = {10.1177/21582440211023178}
}

@article{OliveiraMarcelo2019,
  title = {Investigations into Data Ecosystems: a systematic mapping study},
  volume = {61},
  ISSN = {0219-3116},
  DOI = {10.1007/s10115-018-1323-6},
  number = {2},
  journal = {Knowledge and Information Systems},
  publisher = {Springer Science and Business Media LLC},
  author = {Oliveira,  Marcelo Iury S. and Barros Lima,  Glória de Fátima and Farias Lóscio,  Bernadette},
  year = {2019},
  pages = {589–630}
}

@article{VETRO2016325,
title = {Open data quality measurement framework: Definition and application to Open Government Data},
journal = {Government Information Quarterly},
volume = {33},
number = {2},
pages = {325-337},
year = {2016},
issn = {0740-624X},
doi = {10.1016/j.giq.2016.02.001},
author = {Antonio Vetrò and Lorenzo Canova and Marco Torchiano and Camilo Orozco Minotas and Raimondo Iemma and Federico Morando}
}

@inproceedings{Wohlin2014,
author = {Wohlin, Claes},
title = {Guidelines for snowballing in systematic literature studies and a replication in software engineering},
year = {2014},
isbn = {9781450324762},
publisher = {ACM},
doi = {10.1145/2601248.2601268},
booktitle = {Proceedings of the 18th International Conference on Evaluation and Assessment in Software Engineering},
articleno = {38},
numpages = {10},
address = {London, United Kingdom},
}

@article{Zuiderwijk2014,
author = {Zuiderwijk, Anneke and Janssen, Marijn and Davis, Chris},
year = {2014},
pages = {17-33},
title = {Innovation with open data: Essential elements of open data ecosystems},
volume = {19},
journal = {Information Polity},
doi = {10.3233/IP-140329}
}

@techreport{Trafikverket2023,
  author      = "Trafikverket",
  title       = "Prognos för godstransporter 2040 -Trafikverkets Basprognoser 2023 {(in Swedish)} Forecast for goods transports 2014",
  institution = "Trafikverket",
  year        = "2023",
  type        = "technical report",
  number      = "2023:069",
  month       = "04",
  note        = "https://trafikverket.diva-portal.org/smash/get/diva2:1749263/FULLTEXT01.pdf"
}

@misc{governmentofficesofsweden,
  author       = "{Government Offices of Sweden}",
  title        = "Sweden mobilises to electrify regional freight transport",
  howpublished = "",
  month        = "05",
  year         = "2021",
  note         = "https://www.government.se/articles/2021/05/sweden-mobilises-to-electrify-regional-freight-transport"
}

@article{Nakakojietal2002,
author = {Nakakoji, Kumiyo and Yamamoto, Yasuhiro and Nishinaka, Yoshiyuki and Kishida, Kouichi and Ye, Yunwen},
year = {2002},
pages = {},
title = {Evolution Patterns of Open-Source Software Systems and Communities},
journal = {International Workshop on Principles of Software Evolution (IWPSE)},
doi = {10.1145/512035.512055}
}

@article{Jansen_2014, title={Measuring the health of open source software ecosystems: Beyond the scope of Project Health}, volume={56}, DOI={10.1016/j.infsof.2014.04.006}, number={11}, journal={Information and Software Technology}, author={Jansen, Slinger}, year={2014}, month={Nov}, pages={1508–1519}}

@article{Dulong_de_Rosnay_Janssen_2014, title={Legal and institutional challenges for opening data across public sectors: Towards common policy solutions}, volume={9}, DOI={10.4067/s0718-18762014000300002}, number={3}, journal={Journal of theoretical and applied electronic commerce research}, author={Dulong de Rosnay, Melanie and Janssen, Katleen}, year={2014}, month={Sep}, pages={1–14}}

@article{Rajamäe_Soosaar_Nikiforova_2025, 
title={Bridging the gap: Unravelling local government data sharing barriers in {Estonia} and beyond}, 
volume={56}, 
DOI={10.1016/j.clsr.2024.106099}, 
journal={Computer Law \& Security Review}, 
author={Rajamäe Soosaar, Katrin and Nikiforova, Anastasija}, 
year={2025}, 
month={Apr}, 
pages={106099}}

@misc{EU_GDPR_2016,
  author       = {{European Union}},
  title        = {{Regulation (EU) 2016/679 of the European Parliament and of the Council of 27 April 2016 on the protection of natural persons with regard to the processing of personal data and on the free movement of such data, and repealing Directive 95/46/EC (General Data Protection Regulation)}},
  year         = {2016},
  url          = {https://eur-lex.europa.eu/eli/reg/2016/679/oj?},
  note         = {OJ L 119, 4.5.2016, p. 1–88. Available: 2025-03-17}}

@book{Casestudy2012,
  author       = {Per Runeson and Martin Höst and Austen Rainer and Björn Regnell},  
  title        = {{Case Study Research in Software Engineering: Guidelines and Examples}},  
  year         = {2012},  
  publisher    = {{John Wiley and Sons}},  
  doi          = {10.1002/9781118181034},
  isbn         = {9781118104354},
}

@inbook{RunesonDSSE2020,
	author = {Per Runeson and Emelie Engstr{\"o}m and Margaret-Anne Storey},
	booktitle = {Contemporary Empirical Methods in Software Engineering},
	doi = {10.1007/978-3-030-32489-6\_5},
	editor = {Michael Felderer and Guilherme Horta Travassos},
	pages = {127-147},
	publisher = {Springer},
	title = {The Design Science Paradigm as a Frame for Empirical Software Engineering},
	year = {2020}}

@inproceedings{Larsson_Runeson_Linåker_2024, 
place={Ghent/Leuven, Belgium}, 
booktitle={EGOV-CeDEM-ePart conference}, 
url={https://ceur-ws.org/Vol-3737/paper4.pdf}, 
title={Enabling Data-sharing in Logistics through Open Data Ecosystems – A Literature Review}, 
publisher={EGOV-CeDEM-ePart conference}, 
author={Larsson, John and Runeson, Per and Linåker, Johan}, 
year={2024}}

@book{Trafikverket2025, place={Borlänge}, edition={TRV 2025/37255}, series={ISBN 978-91-8045-477-3}, volume={2025:111}, url={https://trafikverket.diva-portal.org/smash/get/diva2:2002088/FULLTEXT01.pdf}, title={Förslag till nationell plan för transportinfrastrukturen 2026–2037}, publisher={Trafikverket}, author={Trafikverket}, year={2025}, ISBN={ISBN 978-91-8045-477-3}}

@misc{Regeringens_klimathandlingsplan,
  author={Regeringen},
  title={Regeringens klimathandlingsplan – hela vägen
till nettonoll},
  year={2023},
  howpublished= {\url{https://www.regeringen.se/contentassets/990c26a040184c46acc66f89af34437f/232405900webb.pdf}},
  note={Regeringens skrivelse
2023/24:59, Klimat- och näringslivsdepartementet)}
}

@article{PAN_et_al_2021,
title = {Digital interoperability in logistics and supply chain management: state-of-the-art and research avenues towards Physical Internet},
journal = {Computers in Industry},
volume = {128},
pages = {103435},
year = {2021},
issn = {0166-3615},
doi = {https://doi.org/10.1016/j.compind.2021.103435},
author = {Shenle Pan and Damien Trentesaux and Duncan McFarlane and Benoit Montreuil and Eric Ballot and George Q. Huang}
}

@misc{SDGS2023,
  author       = {{United Nations}},
  title        = {Times of Change: Science for Accelerating Transformations to Sustainable Development},
  year         = {2023},
  howpublished = {\url{https://sdgs.un.org/sites/default/files/2023-09/FINAL%20GSDR%202023-Digital%20-110923_1.pdf}},
   note         = {Accessed: 2025-12-01}
}

@techreport{coyle2020valueofdata,
  title        = {The Value of Data: Summary Report},
  author       = {Diane Coyle and Stephanie Diepeveen and Julia Wdowin},
  institution  = {Bennett Institute for Public Policy, University of Cambridge},
  year         = {2020},
  type         = {Technical Report},
  url          = {https://www.bennettschool.cam.ac.uk/wp-content/uploads/2020/12/Value_of_data_summary_report_26_Feb.pdf},
  note         = {Accessed: 2025-12-02}
}

@article{Flyvbjerg2006,
	author = {Bent Flyvbjerg},
	doi = {10.1177/1077800405284363},
	journal = {Qualitative Inquiry},
	number = {2},
	pages = {219-245},
	title = {Five Misunderstandings About Case-Study Research},
	volume = {12},
	year = {2006}}

@INPROCEEDINGS{StrandbergEthical2019,
  author={Strandberg, Per Erik},
  booktitle={ACM/IEEE International Symposium on Empirical Software Engineering and Measurement (ESEM)}, 
  title={Ethical Interviews in Software Engineering}, 
  year={2019},
  volume={},
  number={},
  pages={1-11},
  doi={10.1109/ESEM.2019.8870192}
}

@inproceedings{Moller2020,
  author    = {Frederik M{\"o}ller and Maleen Stachon and Christina Hoffmann and Henrik Bauhaus and Boris Otto},
  title     = {Data-Driven Business Models in Logistics: A Taxonomy of Optimization and Visibility Services},
  booktitle = {Proceedings of the 53rd Hawaii International Conference on System Sciences (HICSS-53)},
  year       = {2020},
  pages      = {5379--5388},
  publisher  = {Hawaii International Conference on System Sciences},
  doi        = {10.24251/HICSS.2020.661},
}

@article{OfeDeReuver2025,
  author  = {Hosea Ofe and Mark de Reuver},
  title   = {Rethinking Openness in Data Platforms: The Impact of Data Artifact Characteristics on Platform Openness Consequences, Scope and Mechanisms},
  journal = {Business \& Information Systems Engineering},
  year    = {2025},
  volume  = {67},
  number  = {6},
  pages   = {959--970},
  doi     = {10.1007/s12599-024-00887-2}
}

@article{Strongetal1997,
author = {Strong, Diane M. and Lee, Yang W. and Wang, Richard Y.},
title = {Data quality in context},
year = {1997},
issue_date = {May 1997},
publisher = {Association for Computing Machinery},
address = {New York, NY, USA},
volume = {40},
number = {5},
issn = {0001-0782},
doi = {10.1145/253769.253804},
journal = {Commun. ACM},
month = may,
pages = {103–110},
numpages = {8}
}

@article{Jussen2024,
  author  = {Ilka Jussen and Frederik M{\"o}ller and Julia Schweihoff and Anna Gie{\ss} and Giulia Giussani and Boris Otto},
  title   = {Issues in Inter-Organizational Data Sharing: Findings from Practice and Research Challenges},
  journal = {Data \& Knowledge Engineering},
  volume  = {150},
  pages   = {102280},
  year    = {2024},
  issn    = {0169-023X},
  doi     = {10.1016/j.datak.2024.102280}
}

@article{Bartolomucci2026,
  author  = {Federico Bartolomucci and Edoardo Ramalli and Valeria Maria Urbano},
  title   = {Integrated Governance in Data Ecosystems: A Conceptual Framework Consolidating Collaborative and Data Governance},
  journal = {Government Information Quarterly},
  volume  = {43},
  number  = {1},
  pages   = {102107},
  year    = {2026},
  issn    = {0740-624X},
  doi     = {10.1016/j.giq.2026.102107}
}

@article{Melville2017,
  author  = {Nigel P. Melville and Terence J. V. Saldanha and Daniel E. Rush},
  title   = {Systems Enabling Low-Carbon Operations: The Salience of Accuracy},
  journal = {Journal of Cleaner Production},
  volume  = {166},
  pages   = {1074--1083},
  year    = {2017},
  issn    = {0959-6526},
  doi     = {10.1016/j.jclepro.2017.08.101}
}

@inproceedings{DalBianco2014,
  author    = {Valentina Dal Bianco and Ville Myllärniemi and Marko Komssi and Mikko Raatikainen},
  title     = {The Role of Platform Boundary Resources in Software Ecosystems: A Case Study},
  booktitle = {2014 IEEE/IFIP Conference on Software Architecture (WICSA)},
  year      = {2014},
  pages     = {11--20},
  publisher = {IEEE},
  doi       = {10.1109/WICSA.2014.41}
}

@article{Jansen2020,
  author  = {Slinger Jansen},
  title   = {A Focus Area Maturity Model for Software Ecosystem Governance},
  journal = {Information and Software Technology},
  volume  = {118},
  pages   = {106219},
  year    = {2020},
  doi     = {10.1016/j.infsof.2019.106219},
}

@article{Adner2017,
  author  = {Ron Adner},
  title   = {Ecosystem as Structure: An Actionable Construct for Strategy},
  journal = {Journal of Management},
  volume  = {43},
  number  = {1},
  pages   = {39--58},
  year    = {2017},
  doi     = {10.1177/0149206316678451},
}

@article{Jacobides2018,
  author  = {Michael G. Jacobides and Carmelo Cennamo and Annabelle Gawer},
  title   = {Towards a Theory of Ecosystems},
  journal = {Strategic Management Journal},
  volume  = {39},
  number  = {8},
  pages   = {2255--2276},
  year    = {2018},
  doi     = {10.1002/smj.2904},
}

@article{RunesonEMSE09,
	author = {Runeson, Per and H\"o{}st, Martin},
	doi = {10.1007/s10664-008-9102-8},
	journal = {Empirical Software Engineering},
	number = {2},
	pages = {131-164},
	publisher = {Springer},
	title = {Guidelines for conducting and reporting case study research in software engineering},
	volume = {14},
	year = {2009}}

@misc{EuropeanCommission2023ESRS,
  author       = {{European Commission}},
  title        = {The Commission Adopts the European Sustainability Reporting Standards},
  year         = {2023},
  date         = {2023-07-31},
  organization = {Directorate-General for Financial Stability, Financial Services and Capital Markets Union},
  url          = {https://finance.ec.europa.eu/news/commission-adopts-european-sustainability-reporting-standards-2023-07-31_en},
  urldate      = {2026-06-14}
}

@article{Olsson2019,
  title = {Framework of Last Mile Logistics Research: A Systematic Review of the Literature},
  volume = {11},
  ISSN = {2071-1050},
  DOI = {10.3390/su11247131},
  number = {24},
  journal = {Sustainability},
  publisher = {MDPI AG},
  author = {Olsson,  John and Hellstr\"{o}m,  Daniel and Pålsson,  Henrik},
  year = {2019},
  month = Dec,
  pages = {7131}
}

@inbook{Sjberg2008,
  title = {Building Theories in Software Engineering},
  ISBN = {9781848000445},
  DOI = {{10.1007/978-1-84800-044-5\_12}},
  booktitle = {Guide to Advanced Empirical Software Engineering},
  publisher = {Springer London},
  author = {Sjøberg,  Dag I. K. and Dybå,  Tore and Anda,  Bente C. D. and Hannay,  Jo E.},
  year = {2008},
  pages = {312–336}
}

@article{Wirtz2022,
  title = {Open government data: A systematic literature review of empirical research},
  volume = {32},
  ISSN = {1422-8890},
  DOI = {10.1007/s12525-022-00582-8},
  number = {4},
  journal = {Electronic Markets},
  publisher = {Springer Science and Business Media LLC},
  author = {Wirtz,  Bernd W. and Weyerer,  Jan C. and Becker,  Marcel and M\"{u}ller,  Wilhelm M.},
  year = {2022},
  month = Sept,
  pages = {2381–2404}
}



\end{document}